\documentclass[aps,prb,twocolumn]{revtex4}
\usepackage{amsfonts}
\usepackage{amsmath}
\usepackage{graphicx}
\usepackage{dsfont}
\usepackage{diagbox}
\usepackage{array}
\usepackage{amssymb}
\usepackage{bbm}
\usepackage{float}
\usepackage{subfigure}
\usepackage{url}
\usepackage{hyperref}
\usepackage{array}
\usepackage{multirow}

\newcommand{\PreserveBackslash}[1]{\let\temp=\\#1\let\\=\temp}
\newcolumntype{C}[1]{>{\PreserveBackslash\centering}p{#1}}
\newcolumntype{R}[1]{>{\PreserveBackslash\raggedleft}p{#1}}
\newcolumntype{L}[1]{>{\PreserveBackslash\raggedright}p{#1}}

\begin{document}

\title{Decoding quantum criticality from fermionic/parafermionic topological
states}
\author{Zi-Qi Wang$^{1}$, Guo-Yi Zhu$^{1}$, and Guang-Ming Zhang$^{1,2}$}
\affiliation{$^{1}$State Key Laboratory of Low-Dimensional Quantum Physics and Department
of Physics, Tsinghua University, Beijing 100084, China. \\
$^{2}$Collaborative Innovation Center of Quantum Matter, Beijing 100084,
China.}
\date{\today}

\begin{abstract}
Under an appropriate symmetric bulk bipartition in a one-dimensional
symmetry protected topological phase with the Affleck-Kennedy-Lieb-Tasaki
matrix product state wave function for the odd integer spin chains, a bulk
critical entanglement spectrum can be obtained, describing the excitation
spectrum of the critical point separating the topological phase from the
trivial phase with the same symmetry. Such a critical point is beyond the
standard Landau-Ginzburg-Wilson paradigm for symmetry breaking phase
transitions. Recently, the framework of matrix product states for
topological phases with Majorana fermions/parafermions has been established.
Here we first generalize these fixed-point matrix product states with the
zero correlation length to the more generic ground-state wave functions with
a finite correlation length for the general one-dimensional interacting
Majorana fermion/parafermion systems. Then we employ the previous method to
decode quantum criticality from the interacting Majorana fermion/parafermion
matrix product states. The obtained quantum critical spectra are described
by the conformal field theories with central charge $c\leq 1$,
characterizing the quantum critical theories separating the
fermionic/parafermionic topological phases from the trivial phases with the
same symmetry.
\end{abstract}

\maketitle

\section{Introduction}

Since the discovery of topological phases of matter, it has been drawn more
and more attention. One of the most prominent signature of topological
phases of matter is the bulk-edge correspondence. For example, the
topological quantum field theory describing the bulk of the fractional
quantum Hall states is in one-to-one correspondence with the conformal field
theory characterizing its edge excitations \cite{Moore_Read}. As a
manifestation of this correspondence, Li and Haldane has shown in a
fractional quantum Hall system that the entanglement spectrum of the bulk
describes the effective low-energy spectrum of the edge physics \cite%
{Haldane_ES_EE_2008}, which was later proved by Qi et. al. in a general
quantum Hall system \cite{Qi_general_ES_top_2012}. In this regards, the
concept of entanglement introduced from quantum information provides a quite
useful tool to characterize the topological phases in condensed matter
systems.

A natural question that follows is, what more we can learn from the
entanglement of the bulk of a topological phase? Is it possible to
encapsulate the information of the quantum critical points adjacent to this
topological phase? At first glance, this may sound crazy because there could
be many paths to drive one phase to another, undergoing different quantum
critical points. However, the phase transition between topological phases
shares some universal features with the gapped topological phase, like the
quantum Hall plateau transition in integer quantum Hall phases is uniquely
determined by the topological number of the adjacent quantum Hall phases. In
fact, in the integer quantum Hall systems and some other symmetry-protected
topological states, the topological phase transition can be viewed as the
result of quantum percolation of the topological edge modes into the bulk
\cite{Chalker}. This implies that via the edge physics, the bulk
entanglement spectrum has the potential to encode information on the
topological quantum criticality. Indeed, it is already verified in the AKLT
states and integer quantum Hall systems \cite%
{Rao_Critical_Entanglement_AKLT,xWan_gmZhang_integer_quantum_Hall_2014,HsiehFu}%
, demonstrating a bulk-edge-criticality correspondence.

But it remains elusive to explore the possibility of decoding topological
quantum criticality from a strongly interacting system with long-range
entanglement. As in the one-dimensional fermionic systems, the interacting
topological phases of matter without protecting symmetry fall into the $Z_{2}
$ classification \cite{Kitaev2010}. The only nontrivial topological phase is
the Kitaev Majorana topological state with hosting an unpaired Majorana zero
mode on the edges. If we go beyond the anomalous free constraint, a natural
generalization of the Majorana topological state is the $\mathbb{Z}_{N}$
parafermion topological state with the defining property of an unpaired
parafermion zero mode on the edges. So it would be interesting to see
whether such a bulk-edge-criticality correspondence is still valid in the
interacting fermionic/parafermionic topological phases via the bulk
entanglement spectrum. Were it be true, this would strongly support the bulk
entanglement study to provide a universal recipe in one-dimensional systems
to extract topological quantum criticality without fine tuning Hamiltonian.

While it is a challenging task to solve a strongly interacting parafermionic
Hamiltonian even in one dimension, the bulk entanglement study has the
privilege that all you need is the ground state wave-function with a finite
correlation length. In general, the matrix product states (MPS) in one
dimension or tensor-network states in two dimension have become a powerful
ansatz to capture the essential feature of the topological ground states. In
particular, the fermionic MPS of the $\mathbb{Z}_{2}$ Majorana topological
phase has been established \cite{Verstraete_fMPS}, which is the minimal
description of the exact ground state of the Kitaev Majorana chain. Our
latest work further generalizes this idea to construct the fixed-point MPS
for the gapped $\mathbb{Z}_{3}$ parafermionic topological phase \cite%
{Xu_parafermionic_MPS} and the more generalized $\mathbb{Z}_{N}$
parafermionic topological MPS. With these new developments, it is much more
efficient to study all kinds of bulk entanglement spectra.

In this paper, we first generalize the fixed-point $\mathbb{Z}_{3}$
parafermionic MPS to a more generic topological $\mathbb{Z}_{N}$
parafermionic MPS with a tunable correlation length. For the wave functions
with a finite correlation length, we show that the topological edge modes
are coupled with the tunable correlation strength. By introducing an
extensive sublattice bipartition, we can derive the bulk entanglement
Hamiltonian that describes a reduced one-dimensional system with
interactions of topological edge modes. Under the symmetric bipartition with
the $\mathbb{Z}_{2}$ fermionic or $\mathbb{Z}_{3}$ parafermionic topological
states in particular, the bulk entanglement Hamiltonians are shown to follow
the behavior of the 1+1 space-time dimensional conformal field theories,
characterizing the topological quantum critical points that separate the
corresponding topological phase from the trivial phase with the same
symmetry. Altogether, we hope that these nontrivial calculations could
evidence a tendency of generalizing the bulk-edge correspondence to the
\textit{bulk-edge-criticality} correspondence.

The rest of the paper will be organized as follows. In Sec. II, we consider
the generic $\mathbb{Z}_{N}$ parafermion topological MPS and discuss its
correlation length. Then we study the single block bipartition for the
parafermionic MPS and its entanglement spectrum which mimics the edge
physics in Sec.III. In Sec. IV, we introduce the procedure of symmetric bulk
bipartition to derive the corresponding reduced density matrix and the bulk
entanglement Hamiltonian. Next we perform exact numerical diagonalization to
the entanglement Hamiltonians and obtain the critical entanglement spectra
for the $\mathbb{Z}_{2}$ Majorana and $\mathbb{Z}_{3}$ parafermionic
topological phases in Sec. V. Finally the discussion of our result and a
brief summary are given in Sec. VI. Some related discussions are included in
the Appendices.

\section{$\mathbb{Z}_{N}$ parafermionic MPS}

In this section we first construct a general $\mathbb{Z}_{N}$ parafermion
topological MPS with a finite correlation length, which is also supported by
the fractionalized Majorana/parafermion zero modes on the edges. Let's start
by introducing the basics of parafermions. In a superconducting system, it
is well-known that the $\mathbb{U}(1)$ charge symmetry of fermions is
usually broken down to the discrete symmetry $\mathbb{Z}_{2}$. Namely, the
particle number conservation is broken down to the number parity
conservation. Causality forbids to break this $\mathbb{Z}_{2}$ parity
symmetry further. Therefore, the fermionic Hilbert space as a super-vector
space can be decomposed into the odd and even parity sectors: $\mathbb{H}=%
\mathbb{H}_{0}\oplus \mathbb{H}_{1}$. Mathematically, this $\mathbb{Z}_{2}$
super-vector space can be generalized to a $\mathbb{Z}_{N}$ super-vector
space $\mathbb{H}=\mathbb{H}_{0}\oplus \mathbb{H}_{1}\oplus \cdots \oplus
\mathbb{H}_{N-1}$. Indeed, the exotic parafermions arising from the
fractionalized topological insulator systems coupled to alternating
ferromagnets and superconductors just live in this super-vector space. The
creation or annihilation operators of $\mathbb{Z}_{N}$ parafermions are the
generalizations of the Majorana fermions that satisfy
\begin{equation}
\chi _{l}^{N}=1,\quad \chi _{l}^{\dagger }=\chi _{l}^{N-1},\quad \chi
_{l}\chi _{l^{\prime }}=e^{i\frac{2\pi }{N}}\chi _{l^{\prime }}\chi _{l},
\label{1.2}
\end{equation}%
for $l<l^{\prime }$. It is easy to check that Majorana fermions fit into the
case of $N=2$. Unlike the bosons, the many-body states of parafermions are
the $\mathbb{Z}_{N}$-graded tensor product of the single-particle states:
\begin{equation}
|k_{1},k_{2},\ldots ,k_{N}\rangle =|k_{1}\rangle \otimes _{g}|k_{2}\rangle
\otimes _{g}\cdots \otimes _{g}|k_{L}\rangle ,
\end{equation}%
where $k$ ranges from $0$ to $N-1$. Exchanging parafermions are
mathematically expressed as an isomorphism for the graded super-vectors as
\begin{eqnarray}
\mathcal{F}\Big(|k_{l}\rangle \otimes _{g}|j_{l^{\prime }}\rangle \Big)
&=&e^{i\frac{2\pi }{N}k_{l}j_{l^{\prime }}}|j_{l^{\prime }}\rangle \otimes
_{g}|k_{l}\rangle ,  \notag \\
\mathcal{F}\Big(\langle k_{l}|\otimes _{g}|j_{l^{\prime }}\rangle \Big)
&=&e^{-i\frac{2\pi }{N}k_{l}j_{l^{\prime }}}|j_{l^{\prime }}\rangle \otimes
_{g}\langle k_{l}|,
\end{eqnarray}%
where $l<l^{\prime }$. An inner product can be defined by mapping the dual
vector space to a complex number i.e. $\mathcal{C}:\mathbb{V}_{F}^{\ast
}\otimes \mathbb{V}_{F}\rightarrow \mathbb{C}$:
\begin{equation}
\mathcal{C}\Big(\langle k_{l}|\otimes _{g}|j_{l^{\prime }}\rangle \Big)%
=\langle k_{l}|j_{l^{\prime }}\rangle =\delta _{k_{l}-j_{l^{\prime }}}.
\end{equation}%
It should be noted that $\mathcal{C}$ and $\mathcal{F}$ commute, which will
be repeatedly used in this paper. Moreover, the way which the parafermion
operator acts on the many-body parafermion state follows as
\begin{eqnarray}
&&\chi _{2j}|\ldots k_{j}\ldots \rangle =e^{i\frac{2\pi }{N}(\sum_{l\leq
j}k_{l}-\frac{N+1}{2})}|\ldots (k_{j}-1)_{\text{mod}~N}\ldots \rangle ,
\notag \\
&&\chi _{2j-1}|\ldots k_{j}\ldots \rangle =e^{i\frac{2\pi }{N}%
\sum_{l<j}k_{l}}|\ldots (k_{j}-1)_{\text{mod}~N}\ldots \rangle ,
\label{ParafOperator}
\end{eqnarray}%
where the phase string arises from the non-local commutation relation of
parafermions. Any parafermionic state should belong to the special
super-vector space with a definite charge detected by the charge operator:
\begin{equation}
\hat{Q}=\prod_{l=1}^{L}(-e^{\frac{i\pi }{N}}\chi _{2l-1}^{\dagger }\chi
_{2l}),~\hat{Q}|\Psi _{m}\rangle =e^{\frac{i2\pi }{N}m}|\Psi _{m}\rangle ,
\end{equation}%
where $|\Psi _{m}\rangle $ is a many-body state defined in the $\mathbb{H}%
_{m}$ with the charge $m=0,1,2,..,N-1$. Recently, the MPS for the
one-dimensional parafermionic topological phase has been constructed \cite%
{Xu_parafermionic_MPS,Xu2}. It can be expressed in terms of a series of
local parafermionic graded tensor:
\begin{equation}
|\Psi _{m}\rangle =\mathcal{C}\left( \mathbf{\hat{\tau}}^{-m}\mathbf{\hat{A}}%
_{1}\otimes _{g}\cdots \otimes _{g}\mathbf{\hat{A}}_{L}\right) .
\end{equation}%
Each local site is associated with a graded product of super-vectors:
\begin{equation}
\mathbf{\hat{A}}\equiv \sum_{\alpha \beta k}A_{\alpha \beta }^{[k]}|\alpha
)\otimes _{g}|k\rangle \otimes _{g}(\beta |,
\end{equation}%
where $|k\rangle $~denotes the Fock parafermion mode and $|\alpha )$~and~$%
(\beta |$~are two virtual parafermion modes living in super-vector space $%
\mathbb{V}_{F}$ and dual-super-vector space $\mathbb{V}_{F}^{\ast }$,
respectively. Contraction of the virtual modes ties the neighboring sites by
maximally entangling bonds, leading to a compact form as
\begin{equation}
|\Psi _{m}\rangle =\sum_{k_{1},\ldots ,k_{L}}\mathcal{C}\mathbf{tr}\left(
\mathbf{\tau }^{-m}A^{[k_{1}]}\cdots A^{[k_{L}]}\right) |k_{1}\cdots
k_{L}\rangle ,  \label{MPS1}
\end{equation}%
with the total charge $m=\sum_{l}k_{l}\mod N$. Here $\mathbf{\tau }$ is the
generator of the $\mathbb{Z}_{N}$ group with its matrix element $\tau
_{\alpha ,\beta }=\delta _{(\beta -\alpha -1)\text{ mod }N}$. For the closed
boundary system, there is a one-to-one correspondence between the total
charge and the boundary conditions. Specifically, in the $\mathbb{Z}_{2}$
Majorana case, $m=0$ corresponds to the even parity state under the
anti-periodic boundary condition, while $m=1$ to the odd parity state under
the periodic boundary condition. A graphic representation for the local
tensor is given in Appendix \ref{RelaPrev}. Motivated by an exact ground
state in the interacting fermionic system \cite{Z2_exact_top_state}, we
generalize the parafermionic MPS to a generic $\mathbb{Z}_{N}$ parafermionic
MPS with a tunable correlation length, and the local tensor is expressed as
\begin{equation}
A_{\alpha ,\beta }^{[k]}=Ce^{-k\phi /N}\delta _{(\beta -\alpha -k)\text{ mod
}N},  \label{LocalTensor}
\end{equation}%
where $\phi $ is a tuning parameter playing the similar role to the chemical
potential for the local charge $k$ ranging from $0$ to $N-1$, and $C$ is the
normalization factor. As shown later, the correlation length can be
continuously tuned by $\phi $. What's amazing is that despite the arbitrary
long correlation length tuned by $\phi $, this MPS always maintains to be
topologically nontrivial and characterizes the topological phase of $\mathbb{%
Z}_{N}$ parafermions away from fixed point. Essentially this is due to the
gauge symmetry
\begin{equation}
\tau ^{\dagger }A^{\left[ k\right] }\tau =A^{\left[ k\right] }.
\end{equation}%
More evidences can be found in the later section. We have to mention that
Eq.~\eqref{LocalTensor} does not exhaust all possibilities of the MPS with $%
N>2$, which maintains the gauge symmetry and topological nontrivial. In fact
for the generic $\mathbb{Z}_{N}$ case, there can be at most $N-1$
independent parameters controlling the relative distribution of the $N$
physical channels on each site $A_{\alpha \beta }^{[k]}\rightarrow
a_{k}\delta _{(\beta -\alpha -k)\text{ mod}~N}$ without breaking the gauge
symmetry. Here and after, we'll abbreviate the modulo $N$ in the arguments
of all delta function for convenience. It should be mentioned that the case
of $\mathbb{Z}_{3}$ is essentially equivalent to the MPS proposed by
Fernando, et. al. \cite{Z3_exact_top_state}. The charge basis we adopt is
nevertheless a better and more natural. Despite its similarity with the
bosonic MPS, the Majorana/parafermionic MPS is dramatically distinct in the
following two aspects: the matrix structure is subjected to the constraint
by the intrinsic $\mathbb{Z}_{N}$ symmetry, and the nontrivial commutation
relation of the (para)fermions brings in a nontrivial phase factor when
performing a contraction or permutation.

Now let's come to discuss the correlation of the general MPS wave function.
A generic two-body correlator $\left\langle \psi \right\vert \hat{O}_{i}\hat{%
O}_{j}\left\vert \psi \right\rangle $ can be cast into the tensor network,
which involves a consecutive mapping through the transfer matrix defined as:
\begin{equation}
\mathbb{E}_{(\alpha \alpha ^{\prime }),(\beta \beta ^{\prime
})}=\sum_{k}A_{\alpha ,\beta }^{[k]}\bar{A}_{\alpha ^{\prime },\beta
^{\prime }}^{[k]}.  \label{TranMat}
\end{equation}%
Note that unlike the bosonic case, the correlator could involve additional
phase factor counting the total charge between the lattice sites $i$ and $j$
due to the parafermion commutation. Nevertheless, this seemingly non-local
phase can be attributed to a local charge detector deposited on the virtual
bonds at the sites $i$ and $j$ only, see Appendix A for detail. According to
a quite standard MPS scheme, the spectrum of this transfer matrix determines
the correlation length of the wave function. To diagonalize this transfer
matrix, we can recast the eigen-equation of the transfer matrix into a
complete positive map:
\begin{eqnarray}
\sum_{k}A^{[k]}R_{n,j}\left( {A}^{[k]}\right) ^{\dagger } &=&\lambda
_{n}R_{n,j},  \notag \\
\sum_{k}\left( A^{[k]}\right) ^{\dagger }L_{n,j}{A}^{[k]} &=&\lambda
_{n}L_{n,j},
\end{eqnarray}%
in which the right eigen-vector is reshaped into the matrix $R_{n,j}$, the
left eigen-vector is reshaped into the matrix $L_{n,j}$, $n$ labels
different eigenvalues, and $j$ labels eigen-vectors with the degenerate
eigen-value. The eigen-equation can be immediately solved as
\begin{eqnarray}
\left( R_{n,j}\right) _{\alpha ,\alpha ^{\prime }} &=&\frac{1}{\sqrt{N}}%
e^{-i2\pi n\alpha /N}\delta _{\alpha -\alpha ^{\prime }-j},  \notag \\
\left( L_{n,j}\right) _{\beta ^{\prime },\beta } &=&\frac{1}{\sqrt{N}}%
e^{i2\pi n\beta /N}\delta _{\beta -\beta ^{\prime }-j},
\end{eqnarray}%
with the eigenvalues $\lambda _{n}=|\lambda _{n}|e^{i\theta _{n}}$, where
\begin{equation}
\left\{
\begin{split}
|\lambda _{n}|& =\left[ 1+\left( \frac{\sin \frac{\pi n}{N}}{\sinh \frac{%
\phi }{N}}\right) ^{2}\right] ^{-\frac{1}{2}}, \\
\theta _{n}& =\tan ^{-1}\frac{\sin \left( \frac{2\pi n}{N}\right) }{\cos
\left( \frac{2\pi n}{N}\right) -e^{2\phi /N}}.
\end{split}%
\right.
\end{equation}%
The global factor $C$ in Eq.~\eqref{LocalTensor} is chosen to make the
largest eigenvalue $\lambda _{0}=1$, ensuring the normalization condition of
the MPS in the thermodynamic limit. It is found that the transfer spectrum
is $N$-fold degenerate so that both $n$ and $j$ ranges from $0$ to $N-1$.
The degeneracy is no surprise because the transfer matrix inherits the gauge
symmetry from the local matrix as a signature of the topological
nontriviality: $(1\otimes \tau ^{\dagger })\mathbb{E}(1\otimes \tau )=%
\mathbb{E}$. The eigenvectors within the degenerate subspace are therefore
related by $(1\otimes \tau )$ or $(\tau \otimes 1)$. And yet $N-1$ among
them, i.e. those non-diagonal ones with $j\neq 0$ are redundant and do not
contribute to the physical consequences. The origin of redundancy is due to
the mismatch between the virtual bond dimension $N$ and the parafermion
quantum dimension $\sqrt{N}$. The physically relevant ones are the diagonal
ones $R_{n,0}$ and $L_{n,0}$. Since $|\lambda _{n}|\leq 1$ and $\lambda _{n}=%
\bar{\lambda}_{N-n}$, the sub-dominant eigenvalue $\lambda _{1}=\bar{\lambda}%
_{N-1}$ contributes to the correlation in the thermodynamic limit:
\begin{equation}
\left\langle \psi \right\vert \hat{O}_{i}\hat{O}_{j}\left\vert \psi
\right\rangle -\left\langle \psi \right\vert \hat{O}_{i}\left\vert \psi
\right\rangle \left\langle \psi \right\vert \hat{O}_{j}\left\vert \psi
\right\rangle \propto {\rm Re}\lambda _{1}^{|j-i|},
\end{equation}%
and the more detail is given in Appendix \ref{TM&CorrFunc}. Therefore the
correlation length $\xi $ can be defined by:
\begin{equation}
\xi ^{-1}=\frac{1}{2}\ln \left[ 1+\left( \frac{\sin \frac{\pi }{N}}{\sinh
\frac{\phi }{N}}\right) ^{2}\right] .  \label{xi}
\end{equation}%
Note that the correlation length is always finite except when $\phi =0$,
concurring with the fixed-point MPS discussed before \cite%
{Xu_parafermionic_MPS}. By varying the parameter $\phi $, we can
continuously tune the correlation length of the topological parafermionic
MPS.

\section{Entanglement spectrum and edge physics}

With the topological wave function, we can perform various bipartitions and
study the corresponding entanglement spectrum to probe the topological
nontriviality therein. Besides the gauge symmetry, the topological
nontriviality manifests much more explicitly on the existence of
fractionalized edge modes. To reach the boundary theory, we can make use of
the bulk-edge correspondence, i.e., studying the single block entanglement
spectrum \cite{Haldane_ES_EE_2008,Qi_general_ES_top_2012}.

\begin{figure}[t]
\includegraphics[width=7.9cm]{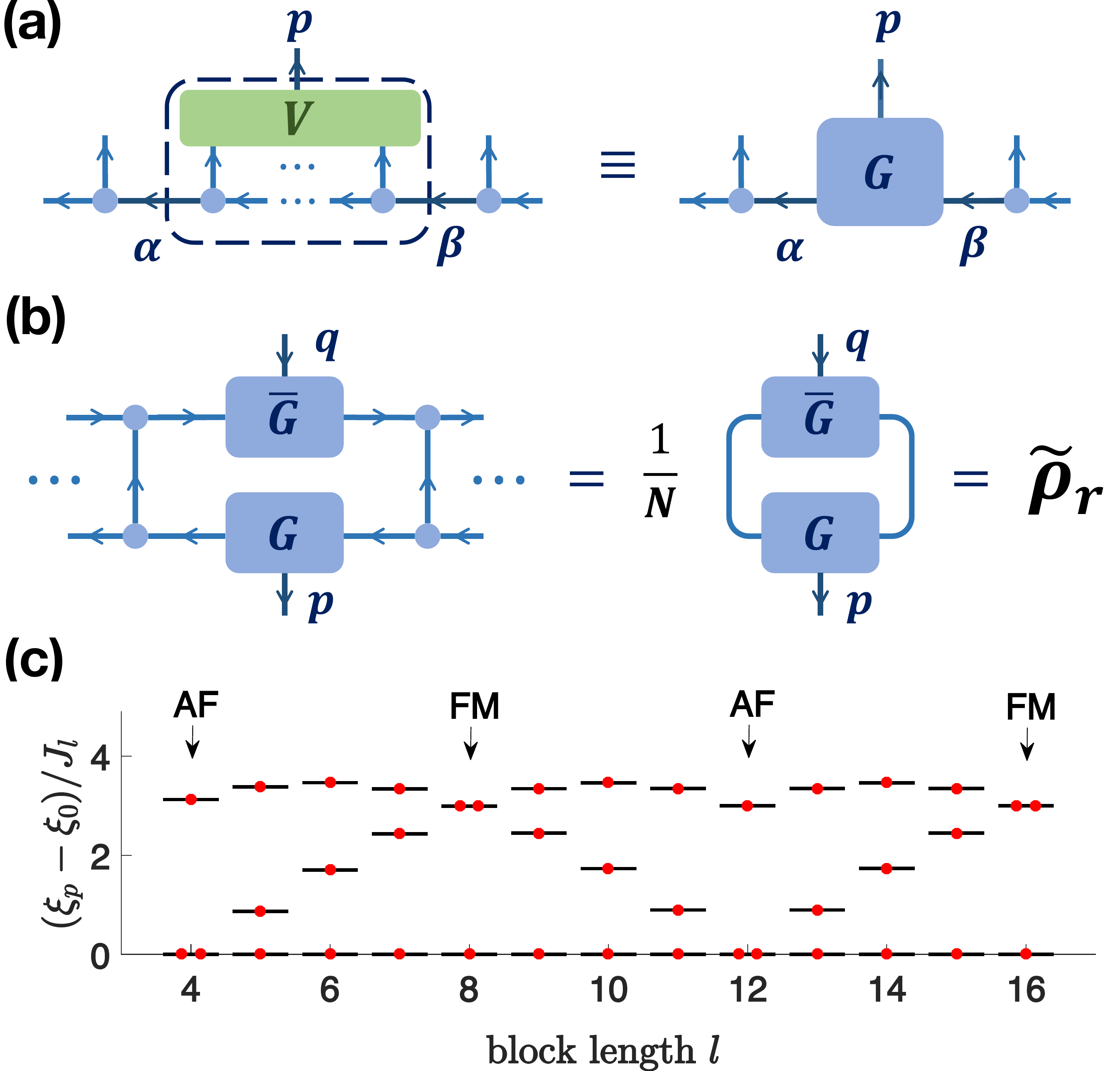}
\caption{(a) Applying an isometry to the block of $l$ sites leads to the
coarse grained MPS. An arrow pointing outwards from one site is associated
to a super-vector space $\mathbb{V}_{F}$ while inwards to $\mathbb{V}%
_{F}^{\ast }$. (b) The reduced density operator of a single block. (c) The
block entanglement spectrum of $\mathbb{Z}_{3}$ case. The spectra structure
experiences a periodicity with respect to the block length. Two typical
phases are remarked as the ferromagnetic (FM) and anti-ferromagnetic (AFM)
between two edge modes. Here we choose $\protect\phi \simeq 1.5076$. }
\label{BlokEnt}
\end{figure}

More specifically, we can bipartite the bulk into one block of $l$ sites and
its complement part, as shown in Fig.~\ref{BlokEnt}(a). By tracing out the
complement part, we're left with a reduced density matrix $\rho _{r}$
describing the block. Since the bulk is gapped, the low energy physics of
the block is reflected on its gapless edge excitations. There is an isometry
transformation $V$ that maps the block physical parafermions $\chi $ to the
effective edge parafermions $\psi $, preserving the spectrum of the
eigenvalues: $\text{spec}(\rho _{r})=\text{spec}(V^{\dagger }\rho _{r}V)$.
In the following we mainly deal with the effective reduced density matrix $%
\tilde{\rho}_{r}\equiv V^{\dagger }\rho _{r}V$, which is supported on the
effective edge degrees of freedom and faithfully characterizes the
low-energy physics of the block. In fact, by treating the complement part as
the environment, we can rewrite the reduced density matrix in terms of the
thermal density matrix as
\begin{equation}
\tilde{\rho}_{r}\equiv e^{-\hat{H}_{\text{ent}}},
\end{equation}%
which defines the entanglement Hamiltonian as the negative logarithm of the
reduced density matrix. The entanglement Hamiltonian was conjectured and
further proved to faithfully characterize the low-energy sectors of the
boundary theory in the quantum Hall systems \cite%
{Haldane_ES_EE_2008,Qi_general_ES_top_2012}. In the following we explain how
to extract the entanglement Hamiltonian using the MPS.

First we coarse grain the MPS by blocking the local matrices of $l$-sites
altogether and treating it as a matrix $\tilde{A}_{\alpha \beta }^{\{k_{i}\}}
$that linearly maps the left- and right- virtual degrees of freedom $(\alpha
\beta )$ into the physical degrees of freedom $\{k_{i}\}$. To extract the
relevant features, we perform a singular value decomposition to the matrix: $%
\tilde{A}=USV^{\dagger }$, where $V$ is exactly the isometry transformation,
$S$ is a diagonal matrix with singular values that characterize the
distributive weights of relevant degrees of freedom, and $U$ as the isometry
that assembles the virtual modes to be the effective degrees of freedom.
According to the relations $\tilde{A}\tilde{A}^{\dagger }=US^{2}U^{\dagger }$
and
\begin{equation}
(\tilde{A}\tilde{A}^{\dagger })_{(\alpha \beta ),(\alpha ^{\prime }\beta
^{\prime })}=\sum_{n,j}\left( R_{n,j}\right) _{\alpha ,\alpha ^{\prime
}}\lambda _{n}^{l}\left( L_{n,j}\right) _{\beta ^{\prime },\beta }.
\end{equation}%
we can deduce $U$ and $S$ by diagonalizing $\tilde{A}\tilde{A}^{\dagger }$.
Therefore $U$ is found with its matrix elements as $U_{(\alpha ,\beta
),p}=\delta _{\beta -\alpha -p}/\sqrt{N}$ and the nonzero singular values in
$S$ are given by
\begin{equation}
s_{p}=\sqrt{\sum_{n=0}^{N-1}\lambda _{n}^{l}e^{i\frac{2\pi }{N}np}},
\label{SingVal}
\end{equation}%
where $p$ ranges from $0$ to $N-1$ and the quantity inside the square root
keeps positively definite. The Eq.\eqref{SingVal} shows how the singular
values rely on the eigenvalue of the transfer matrix and the block length $l$
explicitly. It is worth noticing that, although the two edges together $%
(\alpha \beta )$ appear to have $N^{2}$ degrees of freedom, only $N$ among
them are relevant. This is a signature of the fractionalization, rooted in
the quantum dimension of edge mode being $\sqrt{N}$. Finally we can project $%
\tilde{A}$ onto the relevant degrees of freedom living on edges by the
isometry $G\equiv \tilde{A}V=US$ with its matrix elements
\begin{equation}
G_{\alpha ,\beta }^{[p]}=\frac{s_{p}}{\sqrt{N}}\delta _{\beta -\alpha -p}.
\label{G}
\end{equation}%
When the length of the complement of the block is sufficiently long, shown
in Fig.~\ref{BlokEnt}(b), it is straightforward to obtain $\tilde{\rho}_{r}$
independent of the boundary condition:
\begin{equation}
(\tilde{\rho}_{r})_{p,q}=\mathbf{tr}\left[ G^{[p]}R_{0,0}\left(
G^{[q]}\right) ^{\dagger }L_{0,0}\right] =\frac{1}{N}s_{p}^{2}\delta _{p-q},
\label{eqBlkRDM}
\end{equation}%
where $L_{0,0}$ and $R_{0,0}$ are the matrices reshaped from the
eigenvectors of the transfer matrix corresponding to the maximum eigenvalue.
It is obvious that $\tilde{\rho}_{r}$ is already in a diagonal form with
eigenvalues $s_{p}^{2}/N$. The entanglement spectrum as the eigenvalues of $%
\hat{H}_{\text{ent}}$ is thus given by
\begin{equation}
\xi _{p}=-\ln \left( \frac{1}{N}\sum_{n=0}^{N-1}\lambda _{n}^{l}e^{\frac{%
i2\pi }{N}np}\right) .  \label{BlkEntSpec}
\end{equation}%
From the expression, we find that the complete spectrum decays exponentially
with the block length $l$: $\xi _{p}-\xi _{0}\propto e^{-l/\xi }\equiv J_{l}$%
.

To give a concrete example, we consider the $\mathbb{Z}_{2}$ Majorana
topological state as a parent state. There are only two levels in the block
entanglement spectrum labeled by even and odd parity respectively, and the
gap between them is proportional to $J_{l}$. To understand the spectrum more
transparently, we need the entanglement Hamiltonian that indicates how the
edge degrees of freedom interact with each other. Taking the logarithm of $%
\tilde{\rho}_{r}$ and expressing it back in terms of operators, we can
obtain the entanglement Hamiltonian to the leading order of $J_{l}$:
\begin{equation}
\hat{H}_{\text{ent}}\simeq iJ_{l}\gamma _{1}\gamma _{2}+\ln 2+\mathcal{O}%
(J_{l}^{2}),
\end{equation}%
which describes the coupling between the two edge Majorana zero modes
through the bulk. The coupling strength is exponentially suppressed with
correlation length as the characteristic length. When $l/\xi \gg 1$, the two
edge Majorana zero modes tend to be decoupled, and the excited level tends
to collapse to the ground state level. We then go to a more nontrivial
concrete demonstration with $\mathbb{Z}_{3}$ parafermion topological state.
Similarly, the spectra is exposed to a global exponential suppression due to
the dependence of $J_{l}$ on $l$. But we are more interested in the relative
structure of the entanglement spectrum. The rescaled spectrum for the $%
\mathbb{Z}_{3}$ case is shown in Fig.~\ref{BlokEnt}(c). It is visualized
that the relative structure shows certain periodicity.

Actually, for the generic $\mathbb{Z}_{N}$ cases with $N\geq 3$, there is an
asymptotic periodicity of $2\pi /N$ to the leading order of $J_{l}$:
\begin{equation}
\xi _{p}\simeq \ln N-2J_{l}\cos \left( \frac{2\pi p}{N}+l\theta _{1}\right)
+O(J_{l}^{2}).
\end{equation}%
To understand the spectrum more transparently, the entanglement Hamiltonian
is derived as:
\begin{equation}
\hat{H}_{\text{ent}}\simeq \left[ e^{i(\frac{\pi }{N}+l\theta
_{1})}J_{l}\psi _{1}^{\dagger }\psi _{2}+h.c.\right] +\ln N+O(J_{l}^{2}),
\end{equation}%
which describes the leading coupling between the two effective parafermionic
modes $\psi _{1}$ and $\psi _{2}$ of the edges respectively. Due to the
exponentially decaying of coupling strength, the leading term is sufficient
to characterize the spectrum when block length $l$ is large enough. Now it
is quite clear that the periodicity of the level structure arises from the
coupling phase cycling with the block length. The cases with $l\theta
_{1}=(2k+1)\pi $ and $l\theta _{1}=2k\pi $ are of particular importance. The
former corresponds to the ferromagnetic (FM) coupling of two edge modes in $%
\mathbb{Z}_{3}$ case \cite%
{Tu_critical_Zn_parafermion,Zhuang_Z3_chain_phase_2015}, exhibiting
non-degenerate lowest level and two-fold degenerate excited level, while the
latter corresponds to the anti-ferromagnetic (AFM) coupling of two edge
modes, displaying two-fold degenerate lowest level and non-degenerate
excited level. In the next section we'll show that, depending on the
coupling phase, the distinct interactions lead to distinct quantum critical
theories.

\section{Symmetric bulk bipartition and entanglement Hamiltonian}

While the entanglement Hamiltonian derived from the single block bipartition
is shown to describe the edge physics, it is absolutely incapable of
describing the bulk because it is one-dimension lower than the bulk. In this
section we introduce an extensive sublattice bipartition that yields an
entanglement Hamiltonian describing the bulk physics.

\begin{figure}[t]
\includegraphics[width=7.9cm]{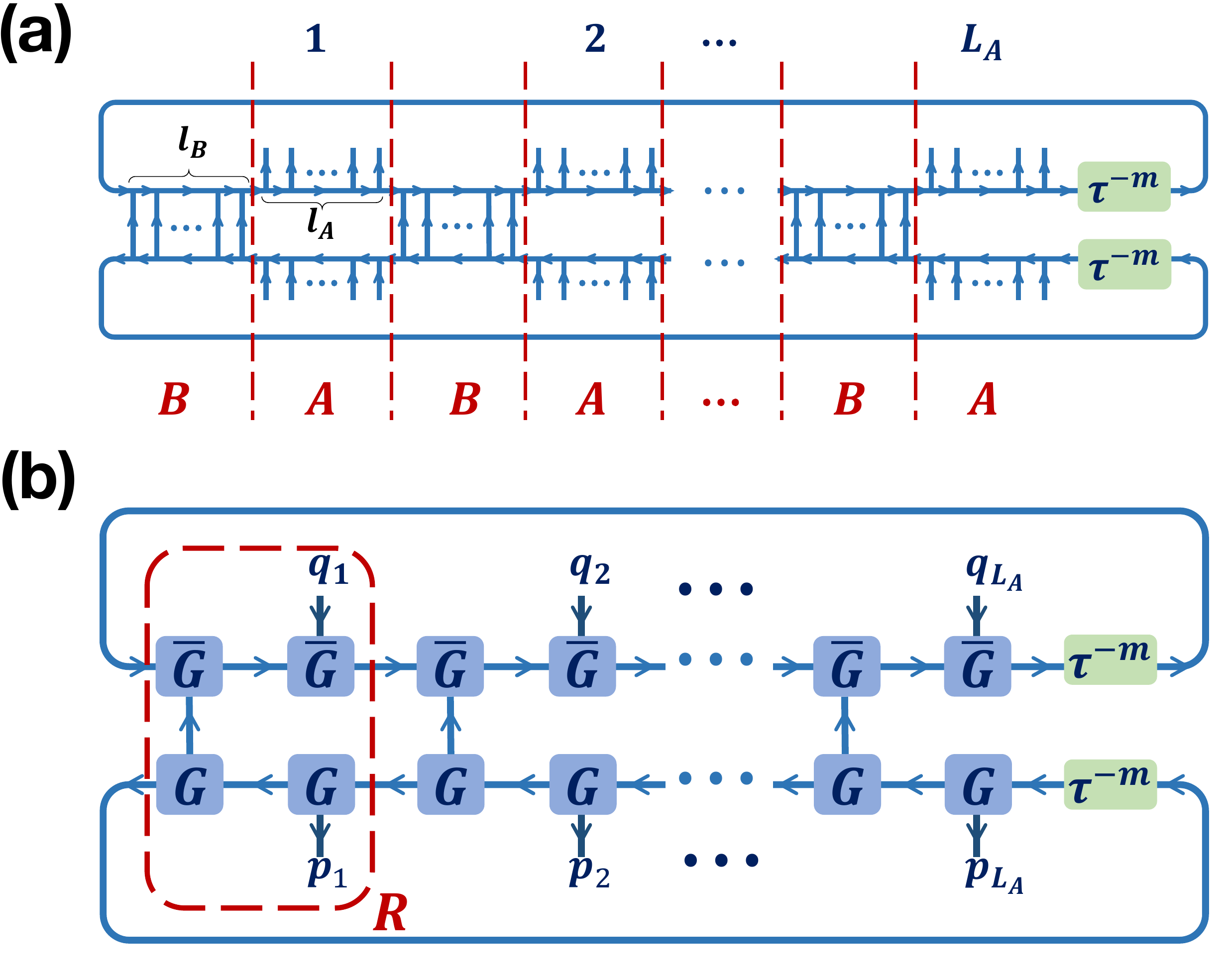}
\caption{(a) A graphical tensor-network representation of the reduced
density operator after a symmetric bulk bipartition. Here both of bra and
ket charge $m$ topological chains are divided into alternating $L_{A}$ pairs
of interlaced A and B parts with respective length $l_{A},l_{B}$. (b) By
coarsed graining both chains, the simplified repeating element is denoted as
R, and the arrows indicate that this tensor-network still carries
parafermionic operators and is not available for the direct numerical
calculation. When all parafermionic states in the part B are contracted, the
reduced density operator becomes a conventional tensor-network, at the prize
of additional bonds playing the similar role of the Jordan-Wigner phase
string.}
\label{SymBulkBiPart}
\end{figure}

The basic idea of decoding topological quantum criticality from a gapped
topological phase \cite{Rao_Critical_Entanglement_AKLT} lies in having the
fractionalized edge degrees of freedom to couple with each other and
percolate into the reduced bulk subsystem. This can be manufactured by a
sublattice bipartition of the bulk into $L_{A}$ pairs of alternating
interlaced A and B sub-blocks, and tracing out the sub-system B and leaving
A as a system of interest. Since the bulk is gapped, the low-energy physics
of subsystem A is dominated by the extensive fractionalized edge degrees of
freedom, as shown in Fig.~\ref{SymBulkBiPart}(a). Distinct from the former
single block bipartition that yields (0+1) dimensional entanglement
Hamiltonian describing the edge physics, this extensive interlaced
bipartition gives rise to a (1+1) dimensional entanglement Hamiltonian
characterizing the bulk properties.

\begin{figure}[t]
\includegraphics[width=7.9cm]{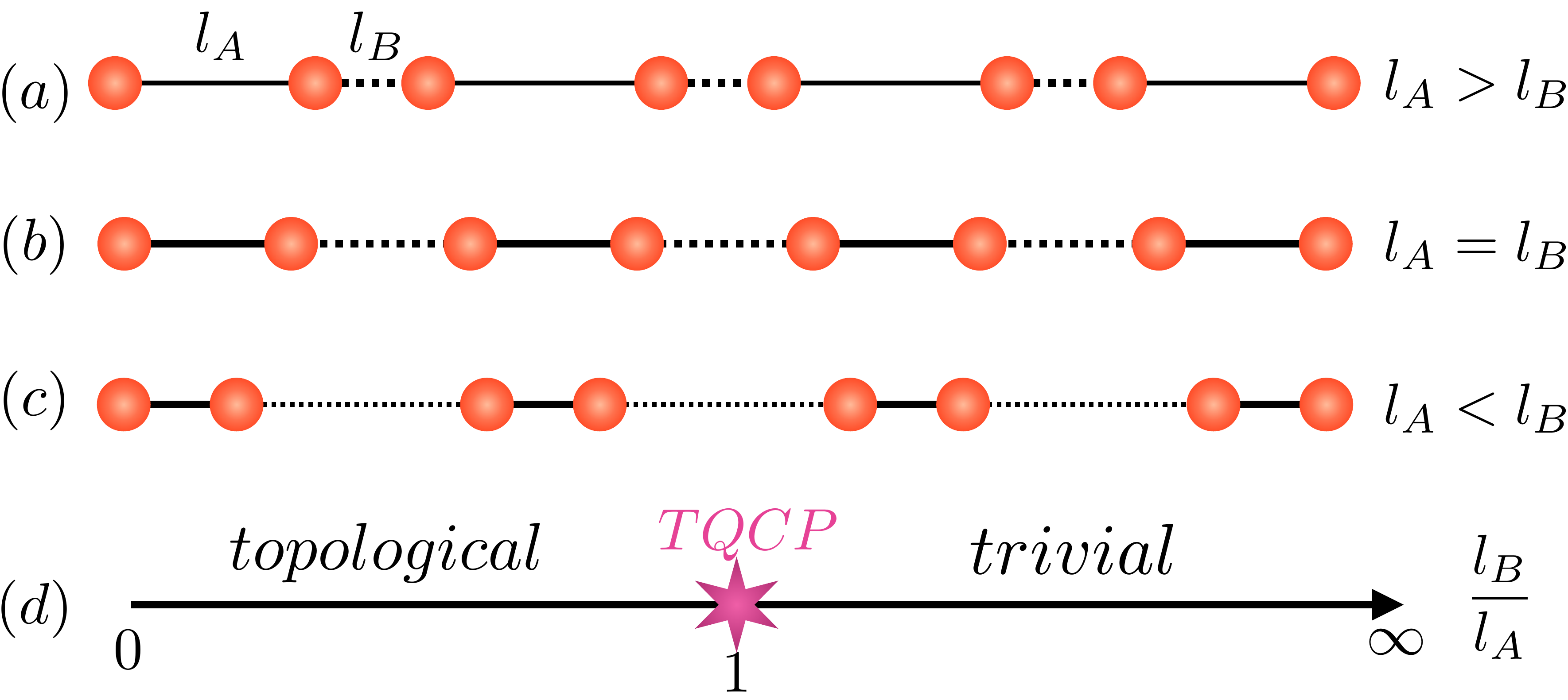}
\caption{(a), (b), (c) The three topologically distinct phases driven by
distinct relative block length are shown schematically. The red dots remark
the fractionalized edge particles as the low energy effective degrees of
freedom of each block in the sub-system A. Solid line corresponds to the
sub-system A while the dashed line stands for the sub-system B that is to be
traced out. The coupling strength between edge particles decay exponentially
with the block length so the relative strong coupling is highlighted with
bold line. (d) A schematic phase diagram. The topological quantum critical
point is guaranteed to be self-dual and translational invariant.}
\label{PhaseDiagram}
\end{figure}

Depending on the relative block length of $l_{A}$ and $l_{B}$, the subsystem
A itself can fall in either topological or trivial phase, as shown in Fig.~%
\ref{PhaseDiagram}. This can be visualized in two limits: when $l_{B}$ is
relatively small, the subsystem A is essentially the same phase with the
original gapped topological phase; when $l_{A}$ is small enough, it falls
into a product state of dimers. As one cannot start from the topological
phase and enter into the trivial phase without experiencing a quantum
critical point, it is expected that the subsystem A could be critical when $%
l_{A}$ and $l_{B}$ are comparable. Especially, in $\mathbb{Z}_{2}$ Majorana
case, the topological phase is related to the trivial phase by a duality
transformation which can be implemented by the single site translation of
the Majorana modes \cite{Alicea_duality_2017}. As a result, the quantum
critical point separating the $\mathbb{Z}_{2}$ Majorana topological phase
from the trivial phase is translational invariant with respect to Majorana
lattice \cite{Hsieh}. This implies that the topological critical point is
robustly pinned at $l_{A}=l_{B}$ despite the presence of even higher order
interactions. This scenario could possibly be generalized to the $\mathbb{Z}%
_{N}$ case, i.e. $l_{A}=l_{B}\equiv l$ to ensure that the coupling strength
between the fractionalized edge modes is translational invariant. We call
this symmetric bulk bipartition.

We first group every $l$ sites together and apply the isometry
transformation $V$ to obtain a coarse grained chain with block tensor $G$.
By tracing out the alternating B blocks, a parafermionic density operator in
the form of a tensor-network is shown in Fig.~\ref{SymBulkBiPart}. However,
the scheme of tracing B is far less straightforward, because it needs
permuting the physical parafermions which would inevitably result in a
highly non-local phase. Nevertheless, this problem can be circumvented by a
trick making use of the tensor-network formalism as a quantum circuit.
Namely, the non-local phase factor arising from commuting parafermions can
be shuffled and redistributed into each site locally at the cost of an
additional bond that tracks and records the total charge of all the
parafermions to the left of certain site. In this way, the parafermionic
density operator turns into a more conventional tensor-network that is
amenable to direct numerical calculation (see more detail in Appendix \ref%
{AddBond}). Moreover, as the parafermion has fractional quantum dimension $%
\sqrt{N}$, the bond dimension of the tensor network can be effectively
reduced from $N^{2}$ to $N$ . Thus the repeating element of the matrix
product operator (MPO) is derived to be a six-rank tensor
\begin{equation}
{R}_{(\alpha l),(\beta r)}^{[p,q]}=\frac{1}{N^{2}}e^{i2\pi (\beta -\alpha -p)%
\frac{l}{N}}\delta _{r-l-p+q}s_{p}s_{q}s_{(\beta -\alpha -p)\text{ mod }%
N}^{2},  \label{R}
\end{equation}%
where $p$ denotes the effective degrees of freedom for each block, $\alpha
,\beta $ are the remained virtual bonds that are to be contracted, and $l$ $%
(r)$ inputs (outputs) the accumulated charge to the left of the site, the
four singular values come from the four blocks respectively, and the phase
factor accounts for the commutation between physical parafermions before
their contraction. It is worth noticing that the additional bonds
essentially play the similar role as the Jordan-Wigner phase string, which
is inevitable when one tries to turn something fermionic into bosonic.

While the reduced density operator in the form of MPO is already available
for direct numerical calculation, we're here to give an analytical
derivation of the final expression of the reduced density operator. In
essence we're going to grain $L_{A}$ number of $R$ tensor together and
contract all the internal bonds, leaving only the physical ones and the
boundary bonds. We denote this intermediate block tensor as $\tilde{R}%
_{\alpha ,\beta r}^{[\{p,q\}]}(L_{A})$, which depends on the length of
subsystem $L_{A}$ and carries both physical and boundary bonds. Finally
contracting the boundary bonds in this block tensor yields the reduced
density matrix, as shown in Fig.~\ref{SymBulkBiPart}(b). By noticing that
the intermediate block tensor has a recursion relation:
\begin{equation}
\tilde{R}_{\alpha ,\beta r}^{[\{p,q\}]}(L_{A})=\sum_{\beta ^{\prime
},r^{\prime }}\tilde{R}_{\alpha ,\beta ^{\prime }r^{\prime
}}^{[\{p,q\}]}(L_{A}-1)R_{\beta ^{\prime }r^{\prime },\beta r}^{\left[
p_{L_{A}}^{~},q_{L_{A}}^{~}\right] },  \label{induction}
\end{equation}%
we can derive the explicit form of the intermediate block tensor, and then
contract the boundary bonds to derive the reduced density operator as the
form
\begin{equation}
\hat{\rho}_{A}=C^{\prime }\left( \prod_{j=1}^{L_{A}}\hat{S}_{j}\right)
\left( \prod_{j=1}^{L_{A}}\hat{I}_{j}\right) \left( \prod_{j=1}^{L_{A}}\hat{S%
}_{j}\right) ,  \label{rho_1}
\end{equation}%
where $C^{\prime }$ is the normalization factor and $\hat{S}$ is an operator
form of the diagonal singular matrix for one block. Physically $\hat{S}$
stands for the coupling between the edge fractionalized particles of one
block, while $\hat{I}$ describes the hopping between fractionalized edge
particles from the adjacent blocks:
\begin{eqnarray}
&&\hat{S}_{j}=\sqrt{\sum_{k=0}^{N-1}\lambda _{k}^{l}\left( -e^{\frac{i\pi }{N%
}}\psi _{2j-1}^{\dagger }\psi _{2j}\right) ^{k}},  \notag  \label{rho_2} \\
&&\hat{I_{j}}={\sum_{k=0}^{N-1}\lambda _{k}^{l}\left( -e^{\frac{i\pi }{N}%
}\psi _{2j}^{\dagger }\psi _{2j+1}\right) ^{k}}.
\end{eqnarray}%
With the reduced density operator, the bulk entanglement Hamiltonian is thus
obtained
\begin{equation}
\hat{H}_{A}\equiv -\text{ln}\hat{\rho}_{A},  \label{BulkEntHam}
\end{equation}%
which is in general a complicated Hamiltonian with long-range interactions
but can be controlled by the correlation length of the parent gapped wave
function.

In the following section we'll numerically perform exact diagonalization for
the reduced density matrix in the form of the MPO, and demonstrate that the
exact numerical results are consistent with the analytical form of the
entanglement Hamiltonian up to the leading orders.

\section{Exact numerical results of critical entanglement spectra}

The main focus of this section is on the bulk entanglement Hamiltonian Eq.%
\eqref{BulkEntHam} defined for the symmetric bulk bipartition. We'll perform
exact diagonalization and obtain a critical spectrum for the entanglement
Hamiltonian labeled by the charge and momentum good quantum numbers. In the
following, we'll first introduce the momentum and charge quantum numbers,
and then show the numerical data for two concrete examples of $\mathbb{Z}%
_{2} $ Majorana phase and $\mathbb{Z}_{3}$ parafermion phase, respectively.
Both of them exhibit quantum critical behavior characterized by conformal
field theories.

\subsection{Symmetries of the entanglement Hamiltonian}

Before doing numerical calculation, we look at the symmetries of the
entanglement Hamiltonian. In fact, the symmetries are inherited from the
parent topological state. The topological states with different charges
correspond to different boundary conditions. As we mentioned early, the $%
\mathbb{Z}_{N}$ charge symmetry $\hat{Q}$ is intrinsic and cannot be broken.
So we always have charge as a good quantum number.

Next we consider the translation symmetry and boundary conditions. As the
symmetry of the entanglement Hamiltonian follow from the parent topological
state, we look into the symmetry actions in the parent state. Since $\mathbf{%
tr}\left( \tau ^{-m}A^{[k_{1}]}\cdots A^{[k_{L}]}\right) =\mathbf{tr}\left(
\tau ^{-m}A^{[k_{L}]}A^{[k_{1}]}\cdots A^{[k_{L-1}]}\right) $ due to the
gauge symmetry, there exists a modified 'bi-translation' symmetry for the $%
\mathbb{Z}_{N}$ topological state with charge-$m$. The bi-translation acts
on the parafermions in the following way:
\begin{equation}
\begin{cases}
\tilde{T}\chi _{j}^{~}\tilde{T}^{\dagger } & =e^{-i\frac{2\pi }{N}}(\chi
_{2}^{\dagger }\chi _{1})^{2}\chi _{j+2}^{~},\quad \text{for }j<2L-1, \\
\tilde{T}\chi _{2L-1}^{~}\tilde{T}^{\dagger } & =e^{i(m-2)\frac{2\pi }{N}%
}(\chi _{2}^{\dagger }\chi _{1})^{2}\chi _{1}^{~}, \\
\tilde{T}\chi _{2L}^{~}\tilde{T}^{\dagger } & =e^{i(m-2)\frac{2\pi }{N}%
}(\chi _{2}^{\dagger }\chi _{1})^{2}\chi _{2}^{~}.%
\end{cases}%
\end{equation}%
Such a bi-translation symmetry physically translates the parafermion lattice
by two sites up to a phase factor depending on the total charge of the
state, and the additional charge zero part $(\chi _{2}^{\dagger }\chi
_{1})^{2}$ plays the role of preserving the parafermion commutation
relation. When $N=2$ for the Kitaev Majorana case, $(\chi _{2}^{\dagger
}\chi _{1})^{2}=-1$, the translation restores a familiar single site
translation, corresponding to the periodic boundary condition for the odd
parity parent state $m=1$, and the anti-periodic boundary condition for the
even parity parent state $m=0$. It is more natural to see that under the
modified translation the physical basis of the generic $\mathbb{Z}_{N}$
parafermion wave-function changes in the following way:
\begin{equation}
\tilde{T}|k_{1},\ldots ,k_{L-1},k_{L}\rangle =|k_{L},k_{1},\ldots
,k_{L-1}\rangle .
\end{equation}%
So the bi-translation symmetry also gives rise to a good quantum number
similar to the usual momentum. The fact that the translation by $L$ times
restores the same operator up to a charge operator is due to the nontrivial
parafermion commutation relation $\tilde{T}^{L}\chi _{j}\tilde{T}^{\dagger
L}=e^{i(m-1)\frac{2\pi }{N}}\hat{Q}^{\dagger 2}\chi _{j}$, leading to the
consequence of shifting the momentum by a fractional value. The boundary
condition of the entanglement Hamiltonian is in one-to-one correspondence
with the charge of the parent topological state. We thus have%
\begin{equation}
\tilde{T}\left( \psi _{j}^{\dagger }\psi _{j+\delta }^{~}\right) ^{k}\tilde{T%
}^{\dagger }=\left( \psi _{j+2}^{\dagger }\psi _{j+2+\delta }^{~}\right)
^{k},  \label{hoping_term}
\end{equation}%
where $\delta $ denotes the neighboring lattice sites and $k$ is a integer
power. The relation $\psi _{2L_{A}+j}^{~}\equiv e^{i(m-1)\frac{2\pi }{N}%
}\psi _{j}$ manifests the boundary condition depending on the charge $m$ of
the parent topological state. Since $\hat{H}_{A}$ is an equal weight
summation of the product of these terms, it can be proved that it is
invariant under the modified "bi-translation" symmetry. Moreover, the charge
operator for those edge modes should be modified as
\begin{equation}
\tilde{Q}=e^{i\frac{2\pi }{N}m}\prod_{j=1}^{L_{A}}(-e^{i\frac{\pi }{N}}\psi
_{2j-1}^{\dagger }\psi _{2j}),
\end{equation}%
where the first factor comes from the charge of the parent topological
state. $\tilde{Q}$ introduces additional charge quantum number $m^{\prime }$
for a given $m$, and such a modified charge operator commutes with the
entanglement Hamiltonian and the modified translation,
\begin{equation}
\left[ \tilde{Q},\hat{H}_{A}\right] =\left[ \tilde{Q},\tilde{T}\right] =0.
\end{equation}%
As a result, we can label the entanglement spectra with the momentum and
charge quantum numbers simultaneously. Last but not least, we mention that
the bi-translation symmetry imposed on the grouped blocks does not depend on
the relative block length. While the momentum is attributed to the
bi-translation symmetry, the single site translation depending on the
relative block length is a much more subtle symmetry.

\subsection{$\mathbb{Z}_{2}$ Majorana fermion criticality}

\begin{figure}[t]
\includegraphics[width=8.5cm]{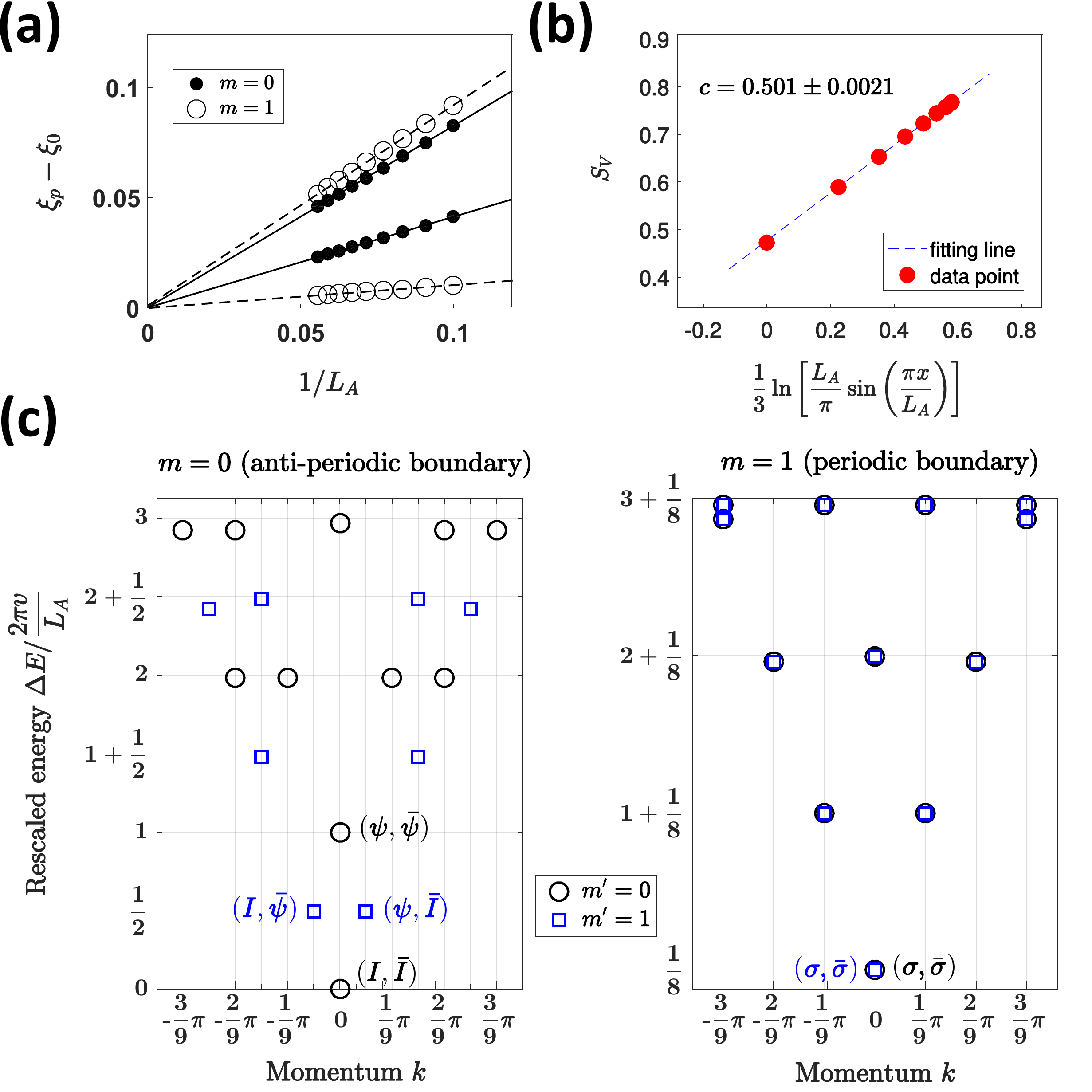}
\caption{(a) The finite-size scaling of the lowest two excited entanglement
levels indicate a critical entanglement spectrum in the thermodynamic limit.
(b) The nested entanglement entropy $S_{V}$ is calculated, and the central
charge is extracted as $c=0.502\pm 0.0021$. (c) The left entanglement
spectrum is obtained from the parent topological state under the
anti-periodic boundary condition ($m=0$), while the right spectrum is
calculated from the parent topological state with the periodic boundary
condition ($m=1$). $m^{\prime }$ denotes the total parity of the
entanglement levels.}
\label{Z2EntSpec}
\end{figure}

As a simple example, we first consider the most familiar $\mathbb{Z}_{2}$
Majorana system. The entanglement spectrum under the symmetric bulk
bipartition is shown in Fig.~\ref{Z2EntSpec}, where we choose $\phi =2$,$%
~l=5 $ and the largest value of $L_{A}=18$. In Fig.~\ref{Z2EntSpec}(a), we
show that the first two energy levels in the finite-size system linearly
collapse to the ground state energy. Meanwhile we can pick out the ground
state of the entanglement Hamiltonian effectively composed of $L_{A}$ sites
and perform the usual block bipartition to calculate the entanglement
entropy. The result is shown in Fig.~\ref{Z2EntSpec}(b), where $x<L_{A}$
denotes the block length. The scaling of entanglement entropy follows the
Calabrese-Cardy formula in the closed boundary \cite{CC_CFT}:
\begin{equation}
S(x,L_{A})=\frac{c}{3}\ln \left[ \frac{L_{A}}{\pi }\sin \left( \frac{\pi x}{%
L_{A}}\right) \right] +S_{0}.
\end{equation}%
The slope yields the central charge $c\simeq \frac{1}{2}$, which uniquely
characterizes the free Majorana fermion CFT. Indeed, it is known that this
theory describes the self-dual critical point separating the $\mathbb{Z}_{2}$
topological phase from trivial phase in Kitaev Majorana chain \cite%
{Alicea_duality_2017,Greiter_Kitaev_to_Ising_2014,chakrabarti_Transverse_Ising}%
.

Moreover, it can be further verified that the low-energy part of the
energy-momentum spectrum exactly follows the scaling law of the Ising
conformal field theory. The energy levels of the lowest primary fields with
their descendants are displayed in Fig.~\ref{Z2EntSpec}(c), where the levels
are shifted by a constant to set the energy of identity primary field as
zero and the values of the levels are rescaled. For the even parity parent
topological state ($m=0$), the boundary condition corresponds to the
anti-periodic in the Majorana fermion representation, and the Hilbert space
of the subsystem A is restricted within the Neveu-Schwarz sector, which
explains why the momentum could be shifted by half-integers in the left
spectrum in Fig.~\ref{Z2EntSpec}(c). The corresponding energy levels are
labeled by the primary fields $(I,\bar{I})$, $(\psi ,\bar{I})$, $(I,\bar{\psi%
})$, $(\psi ,\bar{\psi})$, where the corresponding conformal weights are $%
h_{I}=0$ and$~h_{\psi }=1/2$. On the other hand, for the odd parity parent
topological state ($m=1$), the boundary condition becomes periodic, and the
Hilbert space of subsystem A lies in the Ramond sector. As shown in right
spectrum of Fig.~\ref{Z2EntSpec}(c), the energy levels are marked by the
primary fields $(\sigma ,\bar{\sigma})$ with the conformal weight $h_{\sigma
}=1/16$.

\subsection{$\mathbb{Z}_{3}$ parafermion criticality with FM coupling}

The $\mathbb{Z}_{3}$ parafermion system is more nontrivial because the
existence of parafermions is necessarily a strongly interacting system. In
contrast to the $\mathbb{Z}_{2}$ situation, there could be more than one
quantum critical theories, depending on the phase factor of the coupling
constant between every two neighboring edge parafermions. This phase factor
can be tuned by the block length $l$ in our decoding procedure. Among them
two phases are prototypical: $l\theta _{1}=2k\pi $ for the FM coupling and $%
l\theta _{1}=2(k+1)\pi $ for the AFM coupling.

\begin{figure}[t]
\includegraphics[width=8.5cm]{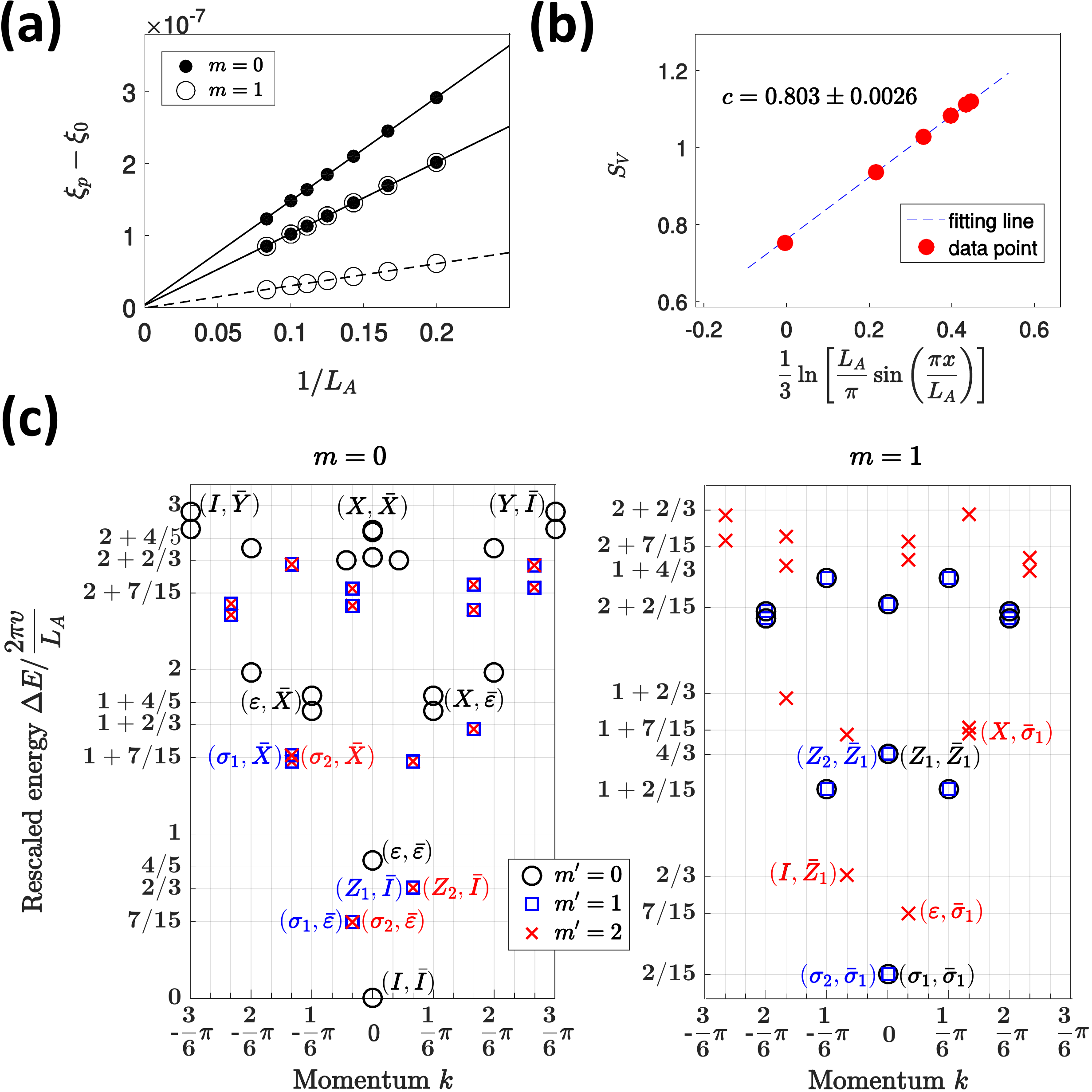}
\caption{(a) The finite-size scaling of the lowest two excited levels under
two different boundary conditions indicate a critical entanglement spectrum
in the thermodynamic limit. (b) The entanglement entropy $S_{V}$ of the
ground state of the entanglement Hamiltonian is calculated and the central
charge is extracted as $c=0.803\pm 0.0026$. (c) The left spectrum is
calculated from the parent topological state with $m=0$, while the right
spectrum is from the parent topological state with $m=1$. $m^{\prime }$
denotes the total charge of the entanglement levels, and the subscript of
the primary fields indicate these total charges.}
\label{Z3FMEntSpec}
\end{figure}

We first consider the FM coupling case. In the numerical calculation, the
parameter is chosen as $l=24$, $\phi \simeq 1.5076$ and the largest values
of $L_{A}=12$. We shall see that the energy levels of the entanglement
Hamiltonian also linearly collapse to the ground state energy, as shown in
Fig.~\ref{Z3FMEntSpec}(a). The entanglement entropy of the ground state of
the entanglement Hamiltonian is also calculated and shown in Fig.~\ref%
{Z3FMEntSpec}(b), which fits into the Calabrese-Cardy formula with a central
charge $c\simeq \frac{4}{5}$, confirming the $\mathbb{Z}_{3}$ parafermion
CFT. Indeed, this theory describes the topological quantum phase transition
from $\mathbb{Z}_{3}$ nontrivial topological phase to the trivial phase.

Furthermore, it can also be shown that the low-energy part of the spectra
exactly follow the scaling law dictated by $\mathbb{Z}_{3}$ parafermion CFT,
displayed in Fig.~\ref{Z3FMEntSpec}(c). Note that the $m=2$ case is related
to $m=1$ by the charge conjugate symmetry and has the same spectrum.
Therefore, we only show the numerical result of $m=0$ and $m=1$. The energy
levels for $m=0$ can be expressed as the conformal primary fields: $\left( I,%
\bar{I}),(I,\bar{Y}),(Y,\bar{I}\right) ,(Y,\bar{Y})$, $(\varepsilon ,\bar{%
\varepsilon}),(\varepsilon ,\bar{X}),(X,\bar{\varepsilon}),(X,\bar{X})$, $%
(\varepsilon ,\bar{\sigma}),(X,\bar{\sigma})$, $(I,\bar{Z})$ and $(Y,\bar{Z})
$. The six corresponding conformal weights are $h_{I}=0$, $h_{\sigma }=1/15$%
, $h_{\epsilon }=2/5$, $h_{Z}=2/3$, $h_{X}=7/5$, and $h_{Y}=3$,
respectively. Meanwhile the energy levels for $m=1$ correspond to the
conformal primary fields: $(\sigma ,\bar{\sigma})$,$~(Z,\bar{Z})$,$%
~(\varepsilon ,\bar{\sigma})$, $(X,\bar{\sigma})$,$~(I,\bar{Z})$ and $(Y,%
\bar{Z})$. The subscripts of the primary fields represent the total
parafermion charge.

\subsection{$\mathbb{Z}_{3}$ parafermion criticality with AFM coupling}

\begin{figure}[tp]
\includegraphics[width=0.48\textwidth]{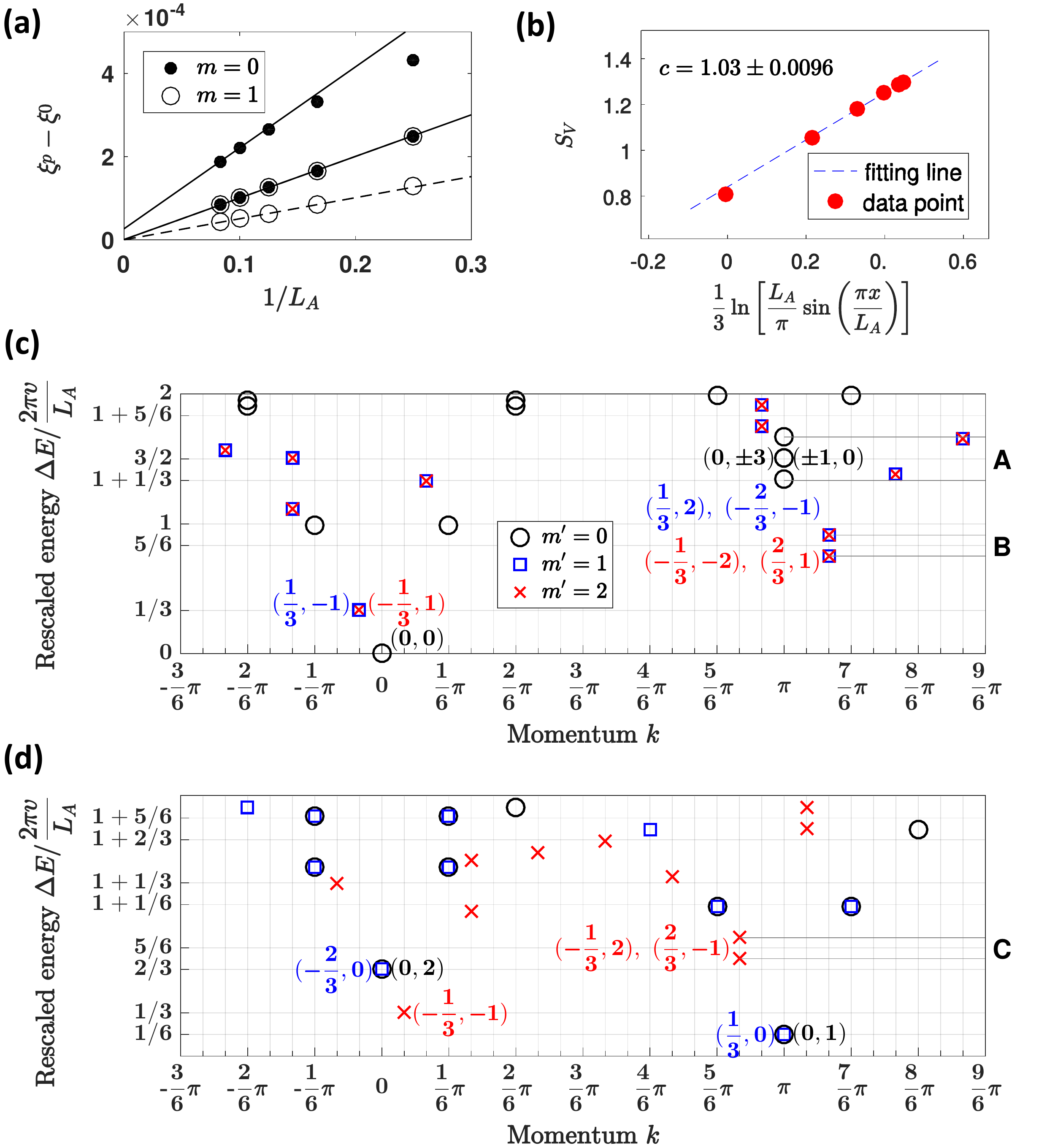}
\caption{(a) The finite-size scaling of the lowest two excited levels under
the two different boundary conditions suggests a gapless entanglement
spectrum in the thermodynamic limit. (b) The entanglement entropy of the
ground state for the entanglement Hamiltonian $S_{V}$ is calculated and the
slope gives the central charge $c=1.03\pm 0.0096$. (c) The entanglement
spectrum is obtained from the $m=0$ parent topological state. (d) The
entanglement spectrum is deduced from the $m=1$ parent topological state.
Some entanglement levels in (c) and (d) have strong finite-size corrections
labeled by the letters A, B, C, whose errors are analysed in Fig.\protect\ref%
{AFerror}. In both (c) and (d), $m^{\prime }$ denotes the total charge of
the entanglement levels labeled by the quantum numbers ($\mathbf{e},\mathbf{m%
}$).}
\label{Z3AFEntSpec}
\end{figure}

Let us now consider the AFM coupling case and perform the similar numerical
calculations for $\phi \simeq 1.5076$, $l=12$ and the largest value of $%
L_{A}=12$. Similar to the FM case, the lowest excitation levels are scaled
linearly as a function of $1/L_{A}$, indicating a gapless spectrum as shown
in Fig.~\ref{Z3AFEntSpec}(a). The calculation of the ground state
entanglement entropy for the entanglement Hamiltonian shown in Fig.~\ref%
{Z3AFEntSpec}(b) determines that the critical entanglement spectrum is
described by the CFT with a central charge $c\simeq 1$.

\begin{figure}[tbp]
\includegraphics[width=0.48\textwidth]{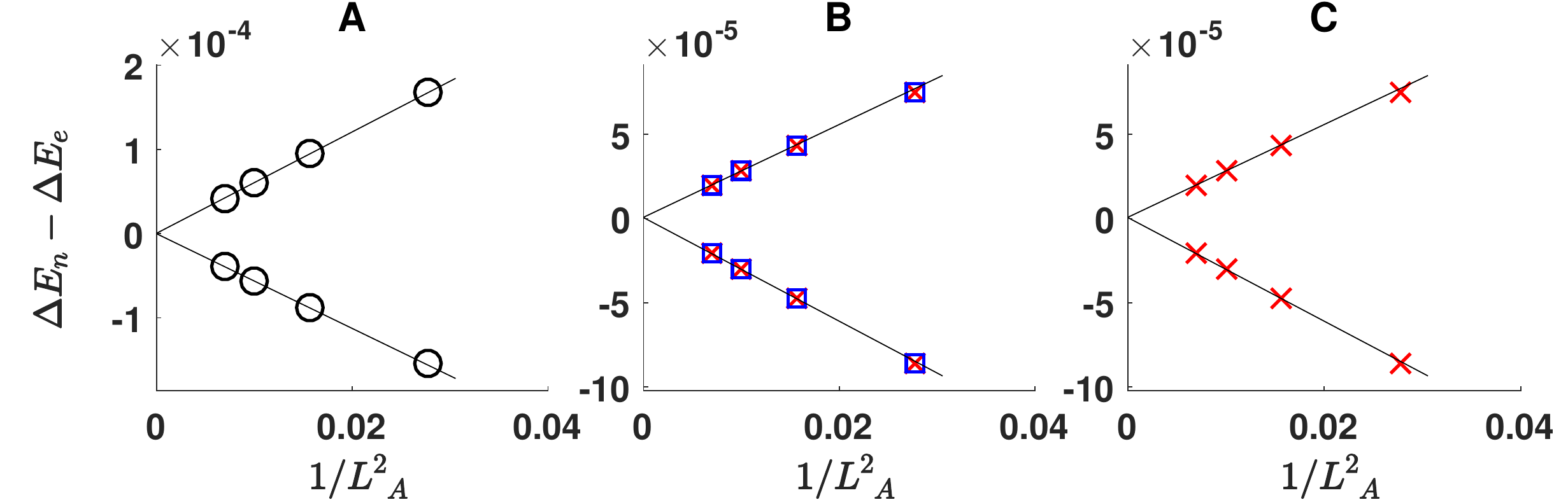}
\caption{The entanglement levels with large finite-size corrections marked
by A, B, C in Fig. \protect\ref{Z3AFEntSpec} (c) and (d) are numerically
analysed, and those corrections are in order of $\mathcal{O}(1/L_{A}^{2})$.}
\label{AFerror}
\end{figure}

Moreover, the low-energy entanglement levels follow the finite-size scaling
law under the three boundary conditions determined by the charge of the
parent topological state $m$, which are displayed in Fig.~\ref{Z3AFEntSpec}%
(c) and (d). Since the large value of $L_{A}=12$, there are finite-size
corrections, which are order of $1/L_{A}^{2}$. In Fig.~\ref{AFerror}, we
have carefully analyzed these corrections so that the entanglement levels
are expressed in terms of the following primary fields: ${(h,\bar{h})}=(0,0)$%
, $(3/4,3/4)$, $(0,1/3)$, and $(3/4,1/12)$ for the charge parent state with $%
m=0$. On the other hand, for the parent topological state with $m=1,2$, the
primary fields include $(1/3,0)$, $(1/12,3/4)$, $(1/12,1/12)$ and $(1/3,1/3)$%
. Note the momenta of some fields have been shifted by $\pi $ in the
presence the AFM correlation. Actually, all the primary fields can be
further represented in terms of the primary fields of the compactified free
boson CFT with the compactified radius $R=\sqrt{2/3}$,%
\begin{equation}
h=\frac{1}{2}\left( \frac{\mathbf{e}}{R}+\frac{\mathbf{m}R}{2}\right) ^{2},~%
\bar{h}=\frac{1}{2}\left( \frac{\mathbf{e}}{R}-\frac{\mathbf{m}R}{2}\right)
^{2},
\end{equation}%
where $\mathbf{e}$ denotes the electrical charge and $\mathbf{m}$ is the
magnetic winding number. In contrast to the critical statistical systems
with both integers of $\mathbf{e}$ and $\mathbf{m}$ values, the quantum
numbers $(\mathbf{e},\mathbf{m})$ are found to be fractional and listed in
Table. \ref{table}.

So the quantum critical point separating a topological phase from its
adjacent trivial phase is not necessarily unique, and we can in practice
deduce a rich family of quantum critical theories to describe the phase
transitions between them.

\begin{table}[tbp]
\caption{The electric charge and winding quantum numbers and their
degeneracies of the primary fields. $m$ stands for the charge of the parent $%
\mathbb{Z}_{3}$ topological states, and $m^{\prime }$ denotes the total
charge of the entanglement levels. }
\label{table}%
\begin{tabular}{ccccc}
\hline\hline
$m$ & $m^{\prime }$ & $(h,\bar{h})$ & $(\mathbf{e},\mathbf{m})$ & degeneracy
\\ \hline
0 & 0 & $(0,0)$ & $(0,0)$ & 1 \\
0 & 0 & $(\frac{3}{4},\frac{3}{4})$ & $(\pm 1,0),~(0,\pm 3)$ & 4 \\
0 & 1 & $(0,\frac{1}{3})$ & $(\frac{1}{3},- 1)$ & 1 \\
0 & 1 & $(\frac{3}{4},\frac{1}{12})$ & $( \frac{1}{3}, 2), ~(-\frac{2}{3},-
1)$ & 2 \\
0 & 2 & $(0,\frac{1}{3})$ & $(-\frac{1}{3}, 1)$ & 1 \\
0 & 2 & $(\frac{3}{4},\frac{1}{12})$ & $(-\frac{1}{3},-2), ~(\frac{2}{3}, 1)$
& 2 \\
1 & 0 & $(\frac{1}{12},\frac{1}{12})$ & $(0, 1)$ & 1 \\
1 & 0 & $(\frac{1}{3},\frac{1}{3})$ & $( 0, 2)$ & 1 \\
1 & 1 & $(\frac{1}{12},\frac{1}{12})$ & $( \frac{1}{3},0)$ & 1 \\
1 & 1 & $(\frac{1}{3},\frac{1}{3})$ & $( -\frac{2}{3},0)$ & 1 \\
1 & 2 & $(\frac{1}{3},0)$ & $(-\frac{1}{3},-1)$ & 1 \\
1 & 2 & $(\frac{1}{12},\frac{3}{4})$ & $(- \frac{1}{3}, 2),~( \frac{2}{3}%
,-1) $ & 2 \\
2 & 0 & $(\frac{1}{12},\frac{1}{12})$ & $(0, -1)$ & 1 \\
2 & 0 & $(\frac{1}{3},\frac{1}{3})$ & $( 0, -2)$ & 1 \\
2 & 1 & $(\frac{1}{3},0)$ & $( \frac{1}{3},1)$ & 1 \\
2 & 1 & $(\frac{1}{12},\frac{3}{4})$ & $(\frac{1}{3},- 2),~( -\frac{2}{3},
1) $ & 2 \\
2 & 2 & $(\frac{1}{12},\frac{1}{12})$ & $(-\frac{1}{3},0)$ & 1 \\
2 & 2 & $(\frac{1}{3},\frac{1}{3})$ & $(\frac{2}{3}, 0)$ & 1 \\ \hline\hline
\end{tabular}%
\end{table}

\section{Discussion and Summary}

To gain a better understanding of the numerical result, we could expand the
logarithm of the reduced density operator Eq.\eqref{rho_1} to obtain the
leading terms of the bulk entanglement Hamiltonian:
\begin{equation}
\hat{H}_{A}\simeq J_{l}\sum_{j=1}^{2L_{A}}\left[ e^{i(\frac{\pi }{N}+l\theta
_{1})}\psi _{j}^{\dagger }\psi _{j+1}+h.c.\right] +L_{A}\ln N,
\label{entangled_H}
\end{equation}%
where the boundary condition is enforced by $\psi _{2L_{A}+1}\equiv e^{i%
\frac{2\pi }{N}(m-1)}\psi _{1}$. Since the couplings between the effective
parafermions are mediated by the gapped bulk and decay exponentially with
the correlation length $J_{l}=e^{-l/\xi }$, the subleading terms to be
relatively negligible can be controlled by tuning $l/\xi \gg 1$. These
leading Hamiltonian terms describe the nearest neighbor hopping of the edge
parafermion modes and can indeed give rise to the quantum critical point
between the topological phase and trivial phase \cite%
{Tu_critical_Zn_parafermion}.

It should be noticed that the quantum critical point that we have extracted
is more than just an isolated singular point, but indeed a phase transition
point on a given path, which is hidden in the decoding operation. As we
already mentioned, by tuning the relative block lengths $l_{A}/l_{B}$ from
larger than $1$ to less than $1$, the subsystem A can undergo a phase
transition from the topological phase to the trivial phase with the same
symmetry, by passing the topological quantum critical point.

In conclusion, we have generalized the recipe of decoding topological
quantum criticality from topological wave-function to the strongly
interacting long-range entangled topological phases. This is achieved by a
nontrivial symmetric extensive interlaced bipartition and extraction of its
entanglement spectrum. This is a novel recipe to extract topological quantum
critical points from a gapped topological ground state, without any need for
the parent Hamiltonian or choosing any specific perturbations. In general,
using this method, we could obtain a family of critical points, for example
the $\mathbb{Z}_{N}$ ($N>2$) parafermion phase. Moreover, we have provided
more than just a method, but our method also has a rather strong implication
of a general physical picture. Our result actually suggests an appealing
generalization of the bulk-edge correspondence to bulk-edge-criticality
correspondence. By some concrete exact demonstrations, we hope that our
results in this paper combined with our earlier works on the bosonic
symmetry-protected states should strongly support that the decoding recipe
and the bulk-edge-criticality are universal in one-dimensional systems. This
could potentially be further generalized to the intrinsic topological order
in two-dimensional systems.

\textit{Acknowledgment.- }The authors would like to thank Wen-Tao Xu and
Hong-Hao Tu for their stimulating discussions and acknowledges the support
of National Key Research and Development Program of China (2017YFA0302902).

\appendix

\section{Relationship between our parafermion MPS to the others\label%
{RelaPrev}}

\begin{figure}[h]
\includegraphics[width=8cm]{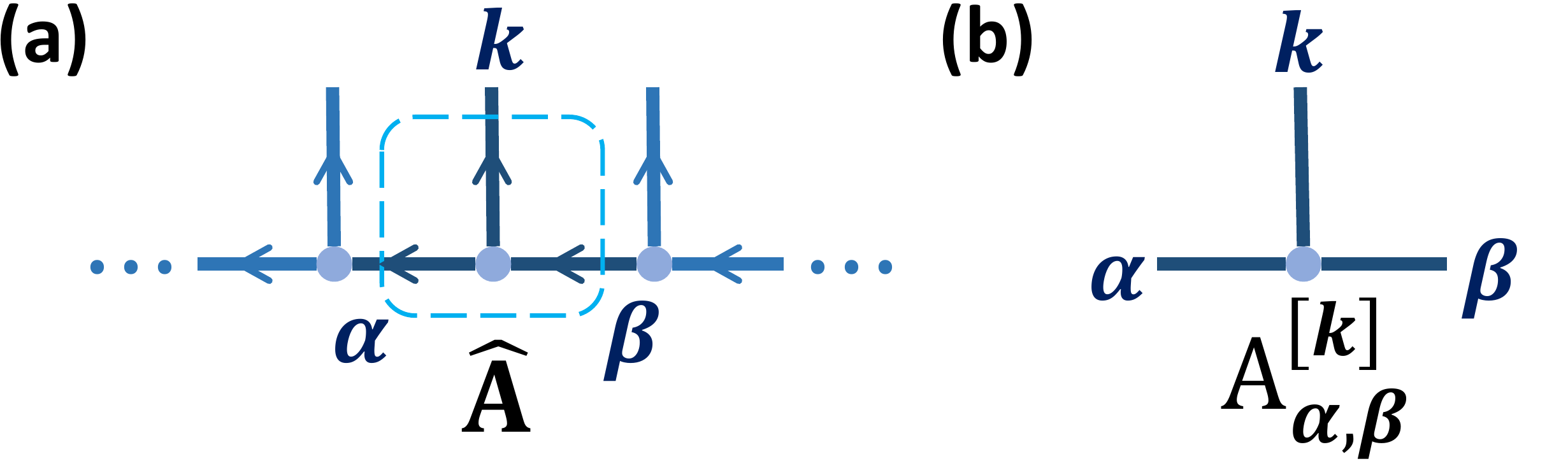}
\caption{(a) The graphical representation of the MPS built from the local
tensor $\hat{A}$. An arrow pointing outwards from one site is associated to
a super-vector space $\mathbb{V}_{F}$ while inwards ones to $\mathbb{V}%
_{F}^{\ast }$. Contraction of the dual vector spaces tie neighboring sites
together. (b) The graphical representation of the coefficient matrix $A^{[k]}
$, where bonds imply the two virtual indices on both sides and one physical
indices on the top. }
\label{MPS}
\end{figure}

Our MPS is defined as the contraction of the local tensors in the graded
space, whose graphic representation is shown in Fig.~\ref{MPS}. It is a
general $\mathbb{Z}_{N}$ parafermionic MPS in the topological phase with the
local matrix:
\begin{equation}
A_{\alpha ,\beta }^{[k]}=Ce^{-k\phi /N}\delta _{\beta -\alpha -k}.
\end{equation}%
For the $\mathbb{Z}_{2}$ case, there is an exact ground state with a finite
correlation length found by H. Katsura, D. Schuricht, and M. Takahashi \cite%
{Z2_exact_top_state}:
\begin{eqnarray}
|\Psi _{0}\rangle  &=&|\Psi ^{+}\rangle +|\Psi ^{-}\rangle ,~|\Psi
_{1}\rangle =|\Psi ^{+}\rangle -|\Psi ^{-}\rangle ,  \notag \\
|\Psi ^{\pm }\rangle  &=&\frac{1}{(1+\alpha ^{2})^{L/2}}e^{\pm \alpha \hat{c}%
_{1}^{\dagger }}e^{\pm \alpha \hat{c}_{2}^{\dagger }}\cdots e^{\pm \alpha
\hat{c}_{L}^{\dagger }}|0\rangle ,
\end{eqnarray}%
where $\hat{c}^{\dagger }$ is the fermionic creation operator, and $\alpha $
is a parameter related to the interaction strength. Because of the Pauli
exclusion principle, all of the exponential terms can be expanded as $1\pm
\alpha \hat{c}_{l}^{\dagger }$, and all the higher order terms vanish
completely, leading to a simple form of these two wave functions:
\begin{eqnarray*}
|\Psi _{0}\rangle  &\propto &\sum_{\{k_{l}\}}|\alpha
|^{\sum\limits_{l=1}^{L}k_{l}}\delta _{(k_{1}+\cdots +k_{L})~\text{mod}%
~2}|k_{1}\ldots k_{L}\rangle , \\
|\Psi _{1}\rangle  &\propto &\sum_{\{k_{l}\}}|\alpha
|^{\sum\limits_{l=1}^{L}k_{l}}\delta _{(k_{1}+\cdots +k_{L}-1)~\text{mod}%
~2}|k_{1}\ldots k_{L}\rangle ,
\end{eqnarray*}%
where the delta function restricts the parity of states to even or odd and
the exponential part has been expressed as the local decaying factors. This
can be further represented into the MPS form with the local tensor:
\begin{equation}
A_{\alpha ,\beta }^{[k]}=e^{-k\phi /2}\delta _{(\beta -\alpha -{k)}~\text{mod%
}~2},~
\end{equation}%
and $\phi =-2\ln |\alpha |$. Now it can be seen that this is nothing but the
$\mathbb{Z}_{2}$ example of our general MPS.

Moreover, another exact solution of~$\mathbb{Z}_{3}$~parafermionic ground
state with a finite correlation length was proposed by F. Iemini, C. Mora,
and L. Mazza \cite{Z3_exact_top_state} in terms of the local tensor
\begin{equation}
A^{[k]}=\left( e^{-\phi /3}\sigma \right) ^{k},~\sigma =%
\begin{pmatrix}
1 & 0 & 0 \\
0 & e^{i\frac{2\pi }{3}} & 0 \\
0 & 0 & e^{i\frac{4\pi }{3}}%
\end{pmatrix}%
.
\end{equation}%
Via a gauge transformation~$U$, we can transform this local tensor into
another charge basis:

\begin{eqnarray}
A^{^{\prime }[k]} &=&U^{\dagger }A^{[k]}U=e^{-k\phi /3}\tau ^{k},  \notag \\
U &=&\frac{1}{\sqrt{3}}%
\begin{pmatrix}
1 & 1 & 1 \\
e^{i\frac{4\pi }{3}} & e^{i\frac{2\pi }{3}} & 1 \\
e^{i\frac{2\pi }{3}} & e^{i\frac{4\pi }{3}} & 1%
\end{pmatrix}%
.
\end{eqnarray}%
Then $A^{\prime }$ is exactly equal to the $\mathbb{Z}_{3}$ specific case of
our general form. Although these two MPS are essentially equivalent to each
other, our charge basis is a better choice. Under the open boundary
condition, our MPS with dangling bond represents a topological state with
zero parafermion modes, while their MPS with dangling bond does not have
such clear physical meaning.

\section{Transfer matrix and correlation functions\label{TM&CorrFunc}}

\begin{figure}[tbp]
\includegraphics[width=8.5cm]{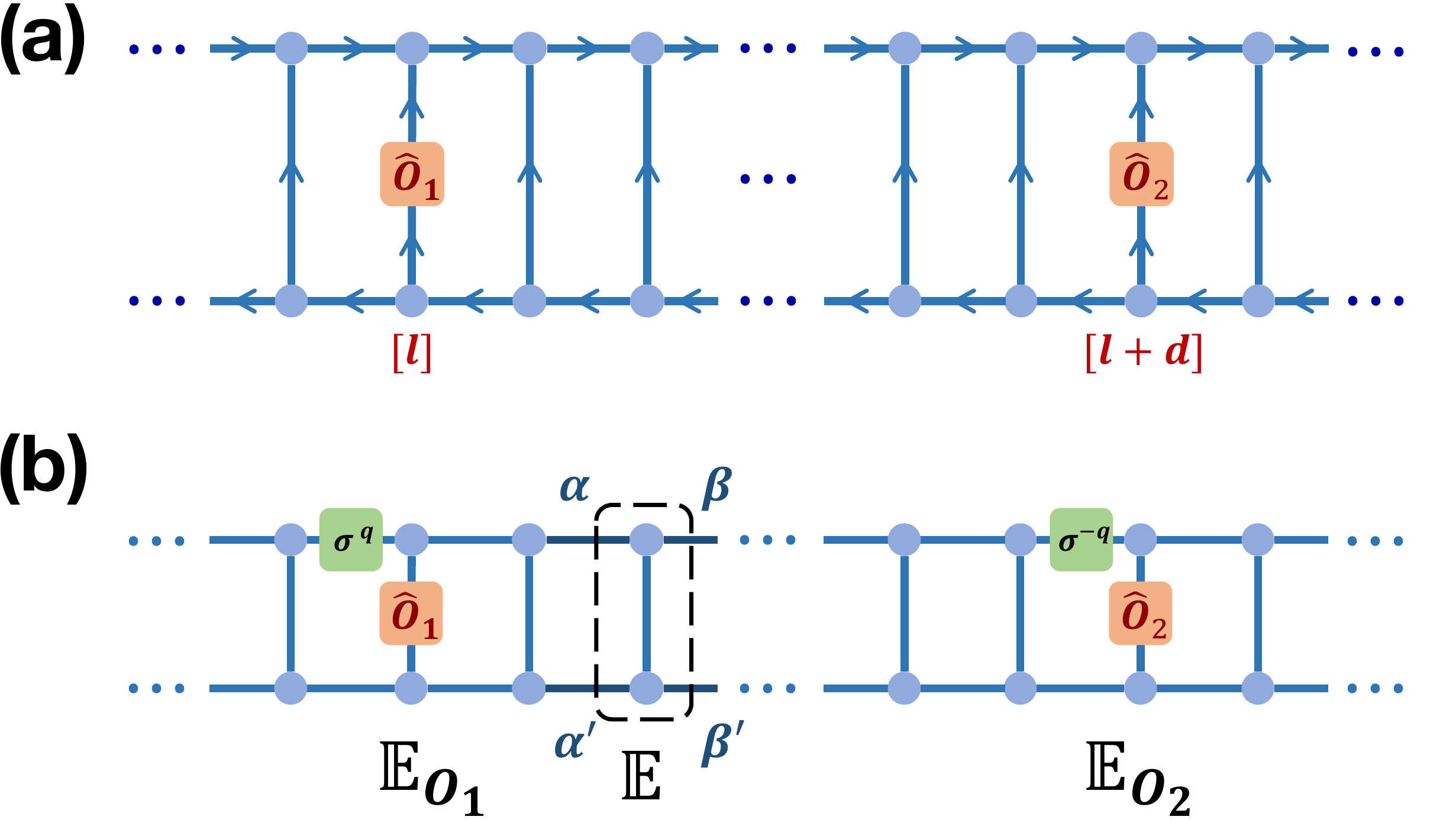}
\caption{{(a)} The definition of a correlation function between the operator
$O_{1}$ with charge $-q$ on the lattice site $l$ and the operator $O_{2}$
with charge $q$ on the lattice site $l+d$ in the chain with length $L$. {(b)%
} The phase factor caused by those contraction can be expressed on the
entangled basis. }
\label{graph_correlation}
\end{figure}

The transfer matrix defined in Eq.\eqref{TranMat} can be expanded by the
eigenvectors:
\begin{equation}
\mathbb{E}_{(\alpha ,\alpha ^{\prime }),(\beta ,\beta ^{\prime
})}=\sum_{n,j}\left( R_{n,j}\right) _{\alpha ,\alpha ^{\prime }}\lambda
_{n}\left( L_{n,j}\right) _{\beta ^{\prime },\beta }.
\end{equation}%
It is very useful to calculate the two-point correlation function, which is
generally defined by two local operators $\hat{O}_{1}$ and $\hat{O}_{2}$
with opposite charges. Assume that $\hat{O}_{1}$ is an operator with charge 
$-q$ on the lattice site $l$, while $\hat{O}_{2}$ is the other one with charge 
$q$ on the lattice site $l+d$, which can be enforced on the local charge $p$ state as:
\begin{equation}
\hat{O}_{1}|p\rangle \propto |(p-q) ~\text{mod}~ N \rangle,       
\hat{O}_{2}|p\rangle \propto |(p+q) ~\text{mod}~ N  \rangle.
\end{equation}                           
Putting them together ensures the zero total charge of the product operator.
The correlation function is defined $\langle \Psi |\hat{O}_{1}\hat{O}%
_{2}|\Psi \rangle $ in a chain with $L$ lattice sites, as shown in Fig.~\ref%
{graph_correlation}(a). No matter what the coefficient of the specific
wave-function is, the contraction of the graded vector can be performed as:
\begin{equation}
\langle k'_{L}|\otimes _{g}\ldots \otimes _{g}\langle k'_{1}|\hat{O}_{1}\hat{O}%
_{2}|k_{1}\rangle \otimes _{g}\ldots \otimes _{g}|k_{L}\rangle .  \label{cf1}
\end{equation}
Then the contractions can be divided into three parts. First, those bra
parafermions in less than $l$ site can be contracted after crossing two
operators with zero total charge. The second is those bra parafermions
from $l$ to $l+d$ sites will be exchanged only with the operators $\hat{O%
}_{2}$ to be contracted, whose charge $q$ leaves a phase factor $e^{-i\frac{%
2\pi }{N}q\left( k_{l}+\cdots k_{l+d}\right) }$. And the remanent
parafermions can be contracted normally without crossing any sites. It is
almost similar to the correlation function for the bosonic case with an
additional non-local phase between two operators. Fortunately, according to
the bulk-edge correspondence, this phase factor can be represented in the
virtual indices $e^{-i\frac{2\pi }{N}(\beta_{l+d-1} -\alpha_{l} )q}$ , which gives rise
to two gauge matrices behind the $d-2$ sites transfer matrix shown in Fig.~%
\ref{graph_correlation}(b).

Finally, the correlation function is written in a very simply form, where a
unitary transform remains in the definition of the correlation length as
shown:
\begin{eqnarray}
&&\left\langle \hat{O}_{1}\hat{O}_{2}\right\rangle =\frac{\left\langle \phi
_{L}\left\vert \mathbb{E}^{l-1}\sigma ^{q} \mathbb{E}_{1}\mathbb{E}%
^{d-2}\sigma ^{-q}\mathbb{E}_{2}\mathbb{E}^{L-d-l}\right\vert \phi
_{R}\right\rangle }{\left\langle \phi _{L}\left\vert \mathbb{E}%
^{L}\right\vert \phi _{R}\right\rangle }  \notag \\
&&\simeq  \left ( \frac{\lambda _{1}}{\lambda _{0}}\right )
^{d}\langle R_{0}|\sigma ^{q}  \mathbb{E}_{O_{1}}|L_{i}\rangle \langle
R_{i}|\sigma ^{-q}\mathbb{E}_{O_{2}}|R_{0}\rangle +h.c. ,
\end{eqnarray}%
where $\sigma _{\alpha ,\beta }=e^{i\frac{2\pi }{N}\alpha }\delta _{\beta
-\alpha }$ is the diagonal $\mathbb{Z}_{N}$ charge matrix and $\mathcal{O}%
(|\lambda _{2}|^{d})$ terms have been ignored. Then the correlation length
can be straightforwardly derived and equal to that for two zero charged
operators, as shown in Eq.\eqref{xi}.

\section{Additional bonds in the parafermion MPO\label{AddBond}}

The reduced density operator we are going to calculate is actually a
parafermionic MPO, the contractions of which brings in nontrivial phase factors arising
from the parafermion commutations. To deal with this, there is a trick to
introduce additional bonds to keep track of the phase factor.

As shown in Fig.~\ref{SymBulkBiPart}, the reduced density matrix can be
obtained by contracting all of grouped physical indices in the subsystem B labeled
by $h,h^{\prime }$, leaving all of the grouped physical indices from the subsystem A
labeled by $p$, $q$. This problem is not simple like the bosonic case, and
all of contractions are carried out from the left as a convention.
\begin{figure}[t]
\includegraphics[width=8.5cm]{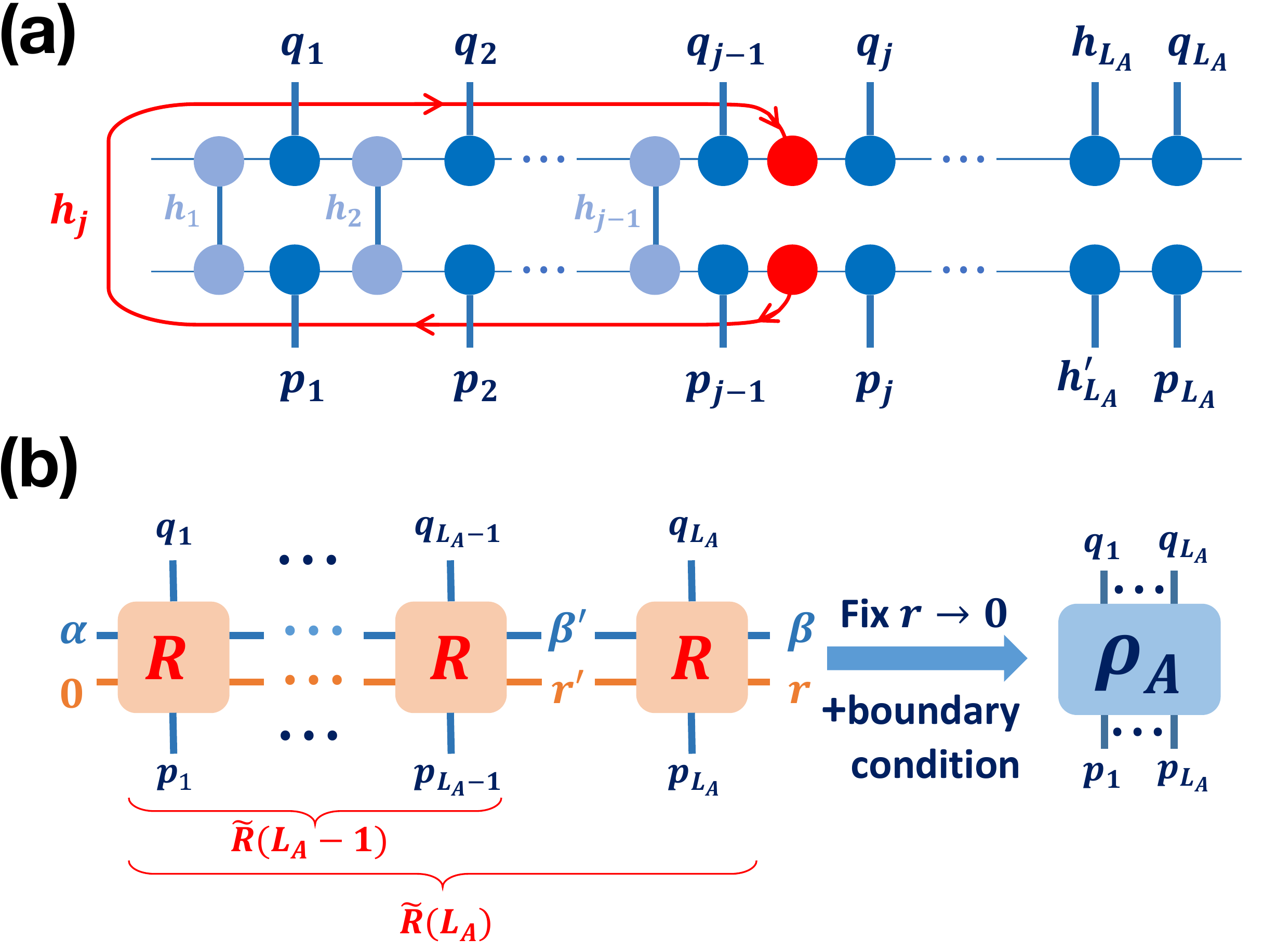}
\caption{(a) The parafermion defined on the $j$-th block of the part B can be
contracted only after crossing $j-1$ parafermions in the part A on both the
bra and ket chains. (b) The reduced density matrix can be obtained by
considering the boundary conditions from the matrix product operator $\tilde{%
R}(L_{A})$, whose unit cell is given by $R$. Those orange lines represent
the additional bonds while the blue ones for the virtual and physical bonds.
}
\label{MPOform}
\end{figure}
Now let's consider about an edge parafermion defined in the graded space of
the $j$-th of the B part labeled by $h_{j}$, it will meet the corresponding
partner labeled by $h_{j}^{^{\prime }}$ only after crossing by the $j-1$
edge parafermions defined in the A part both on bra and ket chains, as shown
in Fig.~\ref{MPOform}(a). Since the directions of exchanges in the bra and ket chains are
opposite, they leave the phase factor conjugated to each other, which
finally causes a phase factor $e^{i\frac{2\pi }{N}h_{j}\sum_{i<j}\left(
p_{i}-q_{i}\right) }$. Those two parafermions labeled by $%
h_{j},h_{j}^{^{\prime }}$ form a bond state with zero charge, so no more
phase factor is present in the next contraction. Here the local index $h_{j}$
represents the virtual indices $\beta _{j}-\alpha _{j}-p_{j}$, therefore,
two additional bonds in the $j$ site are necessary to record the non-local
total charges of parafermions crossed by the parafermions labeled by $h_{j}$, shown as in the $r,l$ bonds of the tensor $%
R $ (Eq.\eqref{R}). This is inspired from the previous work in the fermionic
tensor contractions \cite{extra_bond}. Finally, $r=l+(p-q)$ accumulates the
charge and the phase factor is written into the local form as $e^{i\frac{2\pi}{N} l(\beta_j-\alpha_j-p_j)}$ for the $j$ site.

Further more, the two virtual parafermions living on both of bra and ket
chains constitute a single $N$-dimensional super-vector space, which can be
represented as a single virtual bond. As the result, the reduced density
matrix can be represented into a matrix product operator (MPO), as displayed
in Fig.~\ref{MPOform} with an omitted normalization factor $1/s_{m}(L)$ for
the parent chain with charge $m$ and length $L=2lL_{A}$. The unite cell of the MPO
is defined as $R$:
\begin{equation}
{R}_{(\alpha l),(\beta r)}^{[p,q]}=\frac{1}{N^{2}}e^{i\frac{2\pi }{N}(\beta
-\alpha -p)l}s_{p}s_{q}s_{\beta -\alpha -p}^{2}\delta _{r-l-p+q},  \label{R2}
\end{equation}%
where the left-most bond $l$ is fixed to zero as the start point and the
final accumulated charge (the right-most $r$) is also fixed to zero.

To analyze the reduced density matrix for the subsystem A with $L_{A}$
blocks, the matrix $\tilde{R}(L_{A})$ can be solved mathematically, defined
as Eq. \eqref{induction} and shown in Fig.~\ref{MPOform}(b). After
considering the boundary condition, the reduced density matrix can be
expressed as
\begin{equation}
\left( \rho _{A}\right) _{\{q,p\}}=C^{\prime }\left(
\prod_{j=1}^{L_{A}}s_{p_{j}}\right) \tilde{I}_{\{P,Q\}}\left(
\prod_{j=1}^{L_{A}}s_{q_{j}}\right) .
\end{equation}%
where the singular values are diagonal contributions and $\tilde{I}$
represents the non-diagonal contributions in the middle part of the reduced density matrix. In
the charge accumulating basis, the reduced density matrix can be finally
represented as:
\begin{equation}
\tilde{I}_{\{P,Q\}}=\delta _{P_{L_{A}}-Q_{L_{A}}}\sum_{k=0}^{N-1}\left[ e^{i%
\frac{2\pi }{N}k(m-P_{L_{A}}^{~})}\lambda
_{k}^{l}\prod_{j=1}^{L_{A}-1}\lambda _{k-P_{j}+Q_{j}}^{l}\right] ,
\end{equation}%
where $P_{j}=\sum_{i<j}p_{i}$ $\left( \text{mod }N\right) $ and $%
Q_{j}=\sum_{i<j}q_{i}$ $(\text{mod}~N)$. Moreover, $C^{\prime
}=[N^{L_{A}}s_{m}^{2}(2lL_{A})]^{-1}$ in the expression of $\rho _{A}$ is
the common factor including the normalization factor for the finite length.
From that form, it is easy verified the result given by Eq.\eqref{rho_1},
which is written in the basis of the edge parafermonic operators.

\bibliography{mybibtex}

\end{document}